\documentstyle[11pt,amssymb]{article}
\let\a=\alpha \let\b=\beta \let\g=\gamma \let\d=\delta \let\e=\epsilon
  \let\q=\theta  \let\k=\kappa
\let\l=\lambda \let\m=\mu \let\n=\nu  \let\p=\pi \let\r=\rho
\let\s=\sigma \let\t=\tau    
\let\w=\omega      \let\G=\Gamma   
  \let\S=\Sigma

\let\nn=\nonumber
\def\del{\partial}

\def\R{{{\Bbb R}}}

\def\Z{{{\Bbb Z}}}
\def\B{{{\Bbb B}}}

\newcommand{\be}{\begin{equation}}
\newcommand{\ee}{\end{equation}}
\newcommand{\beq}{\begin{equation}}
\newcommand{\eeq}{\end{equation}}
\newcommand{\ba}{\begin{eqnarray}}
\newcommand{\ea}{\end{eqnarray}}
\newcommand{\bea}{\begin{eqnarray}}
\newcommand{\eea}{\end{eqnarray}}

\def\del{{\partial}}

\def\lbldef#1#2{\expandafter\gdef\csname #1\endcsname {#2}}

\def\href#1#2{#2}

\newcommand{\ber}{\begin{eqnarray}}
\newcommand{\eer}{\end{eqnarray}}
\newcommand{\beqar}{\begin{eqnarray}}

\newcommand{\eeqar}{\end{eqnarray}}

\newcommand{\dsl}{\kern.06em\hbox{\raise.15ex\hbox{$/$}\kern-.56em\hbox{$\partial$}}}

\newcommand{\eeqarr}{\end{eqnarray}}
\newcommand{\ZZ}{{\rm \kern 0.275em Z \kern -0.92em Z}\;}

\begin{document}

\begin{flushright}
\hfill{DAMTP-2002-132}\ \ \
{October 2002}\ \ \
{hep-th/0210218}
\end{flushright}

\vspace{1cm}
\begin{center}
{ \Large {\bf Rotating membranes on $G_2$ manifolds, logarithmic
anomalous dimensions and $N=1$ duality}}

\vspace{1cm}
Sean A. Hartnoll and Carlos Nu\~nez

\vspace{1cm}
{\it DAMTP, Centre for Mathematical Sciences,
Cambridge University\\ Wilberforce Road, Cambridge CB3 OWA, UK}

\vspace{0.5cm}

\noindent S.A.Hartnoll@damtp.cam.ac.uk \hspace{1cm} C.Nunez@damtp.cam.ac.uk
\end{center}
\vspace{1cm}

\begin{abstract}

We show that the $E-S \sim \log S$ behaviour found for long strings rotating
on $AdS_5\times S^5$ may be reproduced by membranes rotating on
$AdS_4\times S^7$ and on a warped $AdS_5$ M-theory solution.
We go on to obtain rotating membrane configurations
with the same $E-K \sim \log K$ relation on $G_2$ holonomy backgrounds
that are dual to ${\mathcal{N}}=1$ gauge theories in four dimensions. We study membrane
configurations on $G_2$ holonomy backgrounds systematically, finding
various other Energy-Charge relations. We end with some comments about
strings rotating on warped backgrounds.

\end{abstract}
\vspace{1cm}


\pagebreak
\setcounter{page}{1}

\tableofcontents
\addtocontents{toc}{\protect\setcounter{tocdepth}{3}}
\vfill\eject

\section{Background and Motivation}

\subsection{A brief history of rotating strings and membranes}

A recent advance in our understanding of the AdS/CFT duality was
the proposal \cite{Gubser:2002tv} that gauge theory operators with large spin
were dual to semiclassical rotating strings in the AdS background.
This original work was
inspired by comments \cite{Polyakov:2001af} concerning `long' gauge theory operators
with high bare dimension and by the success in
matching the anomalous dimensions of large R-charge operators with the
spectrum of string theory on the Penrose limit of $AdS_5\times
S^5$ \cite{Berenstein:2002jq} . String configurations naturally have energies in the
$1/\alpha^{\prime} \sim \lambda^{1/2}$ scale, where $\lambda$ is the 't
Hooft coupling, and are therefore dual to operators with large dimensions.
Rotating strings \cite{Gubser:2002tv} were shown to reproduce
the known results for large R-charge operators \cite{Berenstein:2002jq}, as well as
giving results for a new class of `long' twist two operators.

The principle factor that made the identification of these new
operators possible, was the fact that
a rotating string configuration in AdS space was obtained \cite{Gubser:2002tv}
that had $E - S \sim f(\lambda) \ln S$, 
where $E$ and $S$ are the energy and spin of the configuration. These
must then be dual to gauge theory operators with an 
anomalous dimension that depends logarithmically on the spin. Such
operators were known from the operator approach to Deep Inelastic
Scattering (D.I.S.) in QCD, where they appear in the OPE of electromagnetic
currents. The twist two operators typically have the form
\be\label{eq:theop}
{\mathcal{O}}_S(x) = \Phi(x) \nabla_{\mu_1} \ldots \nabla_{\mu_S} \Phi(0) .
\ee
Where $\Phi(x)$ is a field in the theory such as a field strength or quarks.
The anomalous dimension of these twist two operators is
responsible for violations of Bjorken scaling in D.I.S. at finite coupling
\cite{Peskin}. It is perhaps surprising that the logarithmic
dependence of anomalous dimension on spin survives from the
perturbative to the strong coupling regime in the 't Hooft coupling,
and that no corrections of $\ln^k S$, or other corrections,
appear. Some corrections were shown to vanish in an important one loop
string calculation \cite{Frolov:2002av}. These results
\cite{Frolov:2002av} further clarified the connection with the large
R-charge operators.

The work described so far \cite{Gubser:2002tv} has
subsequently been developed on both the gauge theory and string theory side
in several works \cite{Frolov:2002av,Russo:2002sr,Armoni:2002xp,Armoni:2002fr,Mandal:2002fs,
Zarembo:2002ph,Minahan:2002rc,Axenides:2002zf,Alishahiha:2002sy,Kruczenski:2002fb,
Alishahiha:2002fi,Sezgin:2002rt}. Let us summarise briefly these works.

In \cite{Frolov:2002av,Russo:2002sr}, solutions interpolating smoothly
between the various configurations considered 
in \cite{Gubser:2002tv} were found and a
natural proposal for some of the dual operators was given. In
\cite{Minahan:2002rc}, non-stationary, pulsating, configurations were
considered and, using a WKB approximation, the corrected energies were
found in the limit of large quantum numbers. These configurations were
associated with generalisations of operators with impurities studied
in \cite{Berenstein:2002jq}. In \cite{Mandal:2002fs}, orbifolded
geometries were considered using a collective coordinates approach,
and some connections between gauge theories
and integrable systems were pointed out.
The paper \cite{Armoni:2002xp}
considers strings orbiting around $AdS_5$ 
black holes. Their proposal is to understand the 
finite temperature dual system as a glueball melting into the gluon plasma, due to a 
transfer of angular momentum from the `planetoid' solution 
\cite{deVega:1996mv}, to the black hole.

In \cite{Armoni:2002fr}, an interesting extension was proposed. Firstly, 
they analyse the behaviour expected when one considers 
string configurations in `confining geometries', considering only the
general features of this type of geometry. They propose that 
the functional form has to change, when varying the size of the string 
soliton, from Regge-like ($E \sim S^{1/2}$) to D.I.S-like ($E- 
S \sim \ln S$). A key observation of theirs is that in this case,
unlike the case of $AdS_5\times S^5$, the Regge-like 
behaviour will not be simply an effect of the finite volume in which 
the gauge theory dual is defined. Further, they study string solitons in  
Witten's model for QCD \cite{Witten:1998zw} and find, for a model 
constructed with near extremal Dp-branes, a very curious 
relation of the form $E-S \sim 1 - S^{(p-5)/(9-p)}$. One result of our
work will be to exhibit a Regge to logarithmic transition without
any finite size effects.

In \cite{Alishahiha:2002fi}, the study of strings in Witten's model was
extended by considering pulsating, non-stationary, configurations
similar to those of \cite{Minahan:2002rc}.
Further, \cite{Alishahiha:2002fi} study pulsating membrane
configurations in $AdS_7$ . In \cite{Alishahiha:2002sy}, rotating membranes
were studied in $AdS_7$ spaces, but no logarithmic behaviour for
the anomalous dimensions was found. Instead, power-like behaviours
were displayed. In a very nice paper dealing with higher spin gauge
theories \cite{Sezgin:2002rt}, the authors also discuss membrane
configurations and found similar results. Another result of the
present paper is to obtain a logarithmic anomalous dimension for
membranes in $AdS_4$ (or $AdS_7$).

More recently, the results of \cite{Gubser:2002tv} were reproduced using 
Wilson loops with a cusp anomaly \cite{Kruczenski:2002fb}. 
For the relation of Wilson loops and large R-charge operators, see
for example \cite{Zarembo:2002ph}. Also, the paper
\cite{Axenides:2002zf} has recently studied the anomalous dimensions
on the field theory side.

In this paper, we would like to build on this success by applying the methods of
\cite{Gubser:2002tv} to gauge-gravity dualities that are much less understood than
the canonical $AdS_5\times S^5$ string theory with ${\mathcal{N}}=4$
super Yang-Mills (SYM)
\cite{Maldacena:1997re,Witten:1998qj,Gubser:1998bc,Aharony:1999ti} and
the immediate derivatives thereof. One would like to understand dual descriptions to
${\mathcal{N}}=1$ gauge theories in four dimensions, which have more in common with
observed particle physics. M-theory on a non-compact $G_2$ holonomy manifold is one
way of realising such a gravity dual, as we shall now summarise.

\subsection{Background to the $G_2$ holonomy duality}

Progress in this direction originated from the duality between
Chern-Simons gauge theory on $S^3$ at large $N$ and topological string
theory on a blown up Calabai-Yau conifold
\cite{Gopakumar:1998ki}. This duality was embedded in string theory as
a duality between the IIA string theory of $N$ D6-branes wrapping the
blown up $S^3$ of the deformed conifold and IIA string theory on the
small resolution of the conifold with $N$ units of two form
Ramond-Ramond flux through the blown up $S^2$ and no branes
\cite{Vafa:2000wi}. The D6-brane side of the duality involves an
${\mathcal{N}}=1$ gauge theory in four dimensions that is living on
the non-compact directions of the branes, at energies that do not probe
the wrapped $S^3$.

Before lifting this duality to M-theory, let us make a few further statements regarding
the relation of the D6 branes to the field theory. In
order for the wrapped branes to preserve some supersymmetry, one has to
embed the spin connection of the wrapped cycle into the gauge connection,
which is known as twisting the theory.
On the wrapped part of the brane, the gauge theory is topological \cite{Bershadsky:1995qy}. 
Whilst the twisting allows the configuration to preserve
supersymmetry, some of the supercharges will not have massless
modes. Therefore the theory living on the flat part of the brane will preserve a 
lower fraction of supersymmetry than the unwrapped flat brane configuration.

When we have flat D6 branes, the symmetry group of the 
configuration is $SO(1,6)\times SO(3)_R$. The spinors transform in the
{\bf (8,2)} of the isometry group and the scalars in the
{\bf (1,3)}, whilst the gauge particles are in the {\bf (6,1)}
\cite{Seiberg:1997ax}.
Wrapping the D6 brane on the three-sphere breaks the group to $SO(1,3)\times 
SO(3)\times SO(3)_R$. The technical meaning of twisting is that the 
two $SO(3)$s get mixed to allow the existence of four dimensional
spinors that transform as scalars under the new twisted $SO(3)$. One
can then see that the remaining particles in the 
spectrum that transform as scalars under the twisted $SO(3)$ are the gauge
field and four of the initial sixteen spinors. Thus the field content is that 
of ${\mathcal{N}}=1$ SYM. Apart from these fields, there will be massive modes, whose mass
scale is set by the size of the curved cycle. When we probe the system 
with low enough energies, we find only the spectrum of ${\mathcal{N}}=1$ SYM. In the 
following when we consider `high energies', we will be understanding that the energies are 
not high enough to probe the massive modes of the theory.

The situation is not quite as straightforward as outlined above. This
is because for D6 branes in flat space, the `decoupling' limit does not completely
decouple the gauge theory modes from bulk modes \cite{Itzhaki:1998dd}. In our case, we expect a 
good gauge theory description only when the size of the wrapped three-cycle is 
large, which implies that we have to probe the system with 
very low energies to get 3+1 dimensional SYM \cite{Atiyah:2000zz}. In this case, the size of the two cycle
in the flopped geometry is very near to zero, so a good gravity description 
is not expected. In short, we must keep in mind that the field theory we will be 
dealing with has more degrees of freedom than pure ${\mathcal{N}}=1$ SYM.

It was discovered that the duality described above is naturally
understood by considering
M-theory on a $G_2$ holonomy metric \cite{Atiyah:2000zz}. In eleven
dimensions, $G_2$ holonomy implements ${\mathcal{N}}=1$ as pure gravity.
One starts with a singular $G_2$ manifold that on dimensional
reduction to IIA string theory corresponds to $N$ D6 branes wrapping
the $S^3$ of the deformed conifold. There is an $SU(N)$ gauge theory
at the singular locus/D6 brane. This configuration describes
the UV of the gauge theory. As the coupling runs to the IR, a blown up
$S^3$ in the $G_2$ manifold shrinks and another grows. This flop
is smooth in M-theory physics. The metrics will be discussed in more
detail in the following sections. In the IR regime, the $G_2$
manifold is non-singular and dimensional reduction to IIA gives
precisely the aforementioned small resolution of the conifold with no
branes and RR flux. Confinement emerges
nicely in this picture, because the gauge degrees of freedom have
disappeared in the IR along with the branes. The smooth M-theory
physics of this process was 
systematised in \cite{Atiyah:2001qf} where it was shown that the
transitions are in fact between three possible geometries, corresponding to the
deformed conifold and two small resolutions of the conifold. This
should be understood as a quantum mechanically broken triality
symmetry. See also \cite{Cvetic:2001ya}.
The M-theory lift of the IIA duality of \cite{Vafa:2000wi} was arrived at
independently in \cite{Acharya:2000gb} from the perspective of
studying the M-theory geometry describing four dimensional gauge
theory localised at ADE singularities \cite{Acharya:1998pm}.

The moral of these discoveries would seem to be that special holonomy
in eleven dimensions is a natural way to formulate the dual geometry
of gauge theories living on wrapped D-branes. This approach
was further pursued in, for example,
\cite{Gomis:2001vk,Edelstein:2001pu,Hernandez:2002fb}. The $G_2$
metrics describe the near horizons of branes as opposed to the full brane
supergravity solution because they are not
asymptotically flat. We cannot generically take a further near horizon limit of the
metric, $r\to 0$ typically, because this would spoil the special
holonomy and therefore the matching of supersymmeties. Working within
this paradigm, we shall consider rotating membranes on eleven dimensional
backgrounds $\R^{1,3}\times
X_7$, where $X_7$ is a cohomogeneity-one non-compact $G_2$ manifold.

\subsection{Motivation and contents of this work}

To get going, we will first study rotating membrane
configurations on $AdS_4\times S^7$ and then go on to study rotating membranes on
$G_2$ holonomy manifolds. The first step is something of a warm-up to show that one
obtains sensible results by considering rotating membranes in a configuration that is
fairly well studied. It is however, severely limited by the fact that comparatively little is known about
the dual theory. The second step, on the other hand, is particularly interesting as the duality
takes us to pure ${\mathcal{N}}=1$ SYM theories. This is a theory that is understood and not so
different from the gauge theories of nature. However, what is very poorly understood indeed is the precise
nature of the duality with M-theory on $G_2$ holonomy spacetimes. The
anomalous dimensions of operators with large quantum numbers exhibit very
characteristic behaviours that seem be captured by fairly simple
string/M-theory configurations. It thus provides a window into the duality.

Some rotating membrane configurations on $AdS_7\times S^4$ were
discussed in \cite{Alishahiha:2002sy,Sezgin:2002rt}. 
We will show how a simple modification of their configurations will give logarithms in the
energy-spin relation. This modification will later provide the inspiration for finding logarithms in
the $G_2$ holonomy cases. Another previous use of membrane configurations in $AdS_7\times S^4$
was in providing dual descriptions of Wilson loops in
\cite{Maldacena:1998im}. Also, the presently known matchings of
$N=1$ SYM with $G_2$ holonomy M-theory come from considering membrane
instantons as gauge theory instantons that generate the
superpotential \cite{Acharya:1998pm}, membranes wrapped on one-cycles in the IR
geometry that are super QCD strings in the gauge theory \cite{Acharya:2000gb,Acharya:2001hq}, and fivebranes
wrapped on three-cycles that give domain walls in the gauge theory \cite{Acharya:2000gb,Acharya:2001dz}.
These matchings are essentially
topological and do not use the explicit form of the $G_2$ metrics. In
this sense our results, which do use the explicit form of various
metrics, are of a different nature from previous studies of the duality.

In section 2 we recall the basic formulae for supermembranes and fix
notation. In section 3 we study rotating membranes on AdS spaces that
are dual to gauge theories in three and four dimensions with varying amounts
of supersymmetry. In particular we obtain various configurations with
logarithmic anomalous dimensions. In section 4 we recall the existence of Asymptotically Localy
Conical (ALC) $G_2$ and their role in the ${\mathcal{N}}=1$
duality. We go on to study membranes rotating in these
backgrounds. Again we obtain logarithmic anomalous dimensions, as well
as a variety of other behaviours. In obtaining the logarithms we
consider energy and charge densities of a non-compact brane. Section 6
contains a summary and discussion,
a few comments regarding the dual operators to the membrane
configurations, and open questions. The first appendix is independent from the rest of this work
and sets up a formalism for studying strings moving on warped
backgrounds. The second appendix explicitly checks the lack of
supersymmetry of the $G_2$ holonomy configurations.

\section{Membrane formulae} 

In this section, we briefly summarise the action, equations of motion
and gauge fixing constraints for membranes.
The bosonic sector of the supermembrane action \cite{Bergshoeff:1987qx} is
\be\label{eq:M2action}
{\bf I}_B = \frac{- 1}{(2\p)^2 l_{11}^3} \int d^3\s \left( \frac{(-\g)^{1/2}}{2}
\left[\g^{ij}\frac{\del X^\m}{\del\s^i}\frac{\del X^\n}{\del \s^j}
G_{\m\n}(X) -1 \right] + \e^{ijk} \frac{\del X^\m}{\del\s^i}
\frac{\del X^\n}{\del \s^j}\frac{\del X^\r}{\del \s^k} C_{\m\n\r}(X) \right) ,
\ee
where $i,j,k = 0 \ldots 2$ and $\m,\n,\r = 0\ldots 10$.
The worldsheet metric is $\g_{ij}$, the embedding fields are
$X^{\m}$ and the eleven dimensional background is described by the
spacetime metric $G_{\m\n}$ and three-form field $C_{\m\n\r}$. The
corresponding field strength is $H = dC$.

The equations of motion are
\bea\label{eq:eom}
\g_{ij} & = & \del_i X^\m \del_j X^\n
G_{\m\n}(X) , \nn \\
\del_i \left( (-\g)^{1/2} \g^{ij} \del_j X^\r\right) & = &
- (-\g)^{1/2}\g^{ij} \del_i X^\m \del_j X^\n \G^\r_{\m\n}(X) \nn \\
& & - \e^{ijk} \del_i X^\m \del_j X^\n \del_k X^\s H^\r_{\m\n\s}(X) .
\eea
The action (\ref{eq:M2action}) is equivalent on shell to the
Dirac-Nambu-Goto action. The three diffeomorphism symmetries of the
action may be gauge fixed by imposing the following constraints
\bea\label{eq:constraints}
\g_{0\a} = \del_0 X^\m \del_\a X^\n G_{\m\n}(X) = 0 , \nn \\
\g_{0 0} + L^2 \det \left[ \g_{\a\b} \right] = \del_0 X^\m \del_0 X^\n G_{\m\n}(X) +
L^2 \det \left[ \del_\a X^\m \del_\b X^\n G_{\m\n}(X) \right] = 0 ,
\eea
where $\a,\b=1\ldots 2$ are the spatial worldsheet indices and $L^2$
is an arbitrary constant to be fixed later. Using the equation of motion
for $\g_{ij}$ (\ref{eq:eom}) and the gauge fixing conditions (\ref{eq:constraints}),
one obtains the action
\bea\label{eq:fixed}
{\bf I}_B = \frac{1}{2 (2\p)^2 L l_{11}^3} \int d^3\s \left(
\del_0 X^\m \del_0 X^\n
G_{\m\n}(X) - L^2 \det\left[ \del_\a X^\m \del_\b X^\n G_{\m\n}(X)
\right] \right. \nn \\
\left. + 2 L \e^{ijk} \del_i X^\m
\del_j X^\n \del_k X^\r C_{\m\n\r}(X)
\right) .
\eea
Note that for backgrounds where the $C$ field does not couple to the membrane,
the second constraint in (\ref{eq:constraints}) is just the constancy of the
Hamiltonian of the action (\ref{eq:fixed}).

For the simple
configurations we consider below, that have additional conserved
charges, the equations of motion will almost follow from imposing the constant
Hamiltonian constraint. However, in making ans\"atze for solutions one needs to be
quite careful about consistency. Warped terms are particularly
dangerous. The equations of motion for the gauge fixed action
(\ref{eq:fixed}) contain terms like $\del_0 X^{\m} \del_0 X^{\n}
\del_{\r} G_{\m \n}(X)$, and these generally have to vanish in order
for the equations of motion to be solved. This must be checked for each
anstaz adopted.

\section{Membranes rotating in AdS geometries}

This section considers membranes rotating in various AdS
backgrounds. These configurations are very straightforward
generalisations of previous work and we consider this section to be a
warm-up for the $G_2$ cases to be considered below. We modify previous
configurations slightly to obtain logarithmic terms in energy-spin relations.
We call these new
configurations Type I and the previously studied, non-logarithmic, membrane
configurations Type II. We emphasise that this distinction, and the
existence of logarithms, is independent of the precise AdS
geometry, so long as the internal manifold has a $U(1)$ isometry.

\subsection{Membranes rotating in $AdS_4\times M_7$}

We start by studying membranes moving in $AdS_4 \times
M_7$. We will take first the maximally supersymmetric case with $M_7=
S^7$ and then move on to more interesting 
geometries preserving ${\mathcal{N}}=1,2,3$
supersymmetries in the dual 2+1 dimensional theory. 
The dual field theories will be conformal and are, in some aspects, very well 
known. We will study two different types of configurations. 
The first type of configurations, type I, are
similar to the original string configurations \cite{Gubser:2002tv},
and will give logarithmic anomalous dimensions.
Type II configurations are essentially the membrane configurations
that have already been studied \cite{Alishahiha:2002sy} .

The metric and three-form potential are
\beq
\frac{1}{l_{11}^2} ds_{11}^2=- \cosh^2\rho dt^2 + 
d\rho^2 + \sinh^2\rho (d\alpha^2 + \sin^2\alpha d\beta^2) + B^2 
ds_7^2 ,
\label{metrica}
\eeq
\beq
H = k \cosh\rho \sinh^2\rho \sin\alpha dt \wedge d\rho \wedge d\alpha
\wedge d\beta,\;\;\; C= - \frac{k 
\sin\alpha}{3} \sinh^3 \rho dt \wedge d\alpha\wedge d\beta .
\label{gaugefield}
\eeq
Here $B$ is the relative radius of 
$AdS_4$ with respect to the seven-manifold, whilst $k$ is a number 
that can be easily determined from the equations of motion.

Let us first study the case in which $M_7=S^7$. 
In this case we find it convenient to write the metric as
\beq
ds_7^2= 4 d\xi^2 + \cos^2\xi (d\theta^2 + d\phi^2 + d\psi^2 + 2 \cos\theta 
d\phi d\psi) + \sin^2\xi (d\tilde{\theta}^2 + d\tilde{\phi}^2 + 
d\tilde{\psi} + 2 \cos\tilde{\theta}
d\tilde{\phi} d\tilde{\psi}) .
\eeq
We could equally well be considering the case of 
the squashed seven sphere, $\tilde{S}^7$, the supergravity system will 
be dual to a conformal gauge theory with ${\mathcal{N}}=1$ 
supersymmetry. In this case, the metric will read
\beq
ds_7^2= d\Omega_4^2 + \frac{1}{5} (\omega^i - A^i)^2 , 
\label{metricasquashed}
\eeq
with $A^i$ being the SU(2) one-instanton on $S^4$ and $\omega^i$ the
left-invariant one-forms of $SU(2)$ (for details see for example 
\cite{Gursoy:2002tx}). We will see that we obtain the same results in
both cases.

The two types of solution mentioned above differ in
the dependence of the AdS coordinates on the 
worldvolume of the membrane. The membrane is moving forward trivially
in time, one direction is stretched along the radial direction of AdS
and is rotating either in the AdS (spin) or in the
internal, $M_7$, space (R-charge). There is one extra direction left, with
worldvolume coordinate $\d$, that distinguishes the membrane from a
string. We must wrap this direction along a $U(1)$ isometry. This
could either be in the AdS (type I configuration), or in the internal
space (type II configuration).
Thus, for the type I solutions the wrapped direction
of the membrane, $\d$-direction, remains finite at infinity, and the
long membrane limit is string-like. For the type II solutions, the
wrapped direction is not stabilised asymptotically. This kind of
distinction will play an important role below when we discuss
membranes on $G_2$ manifolds.

In constracting these solutions, it is important to check that the
ans\"atze are in fact consistent. In practice, this constrains the
values that one may give to constant angular coordinates.

\subsection{Solutions of type I: logarithms}

As this is our first configuration, let us describe it clearly.
We want to embed the membrane into spacetime such that it
is moving trivially forward in time and is extended along the radial
direction of AdS
\be
t= \kappa \tau,\;\; \rho= \rho(\sigma) .
\label{solution1}
\ee
We would then like to have the membrane rotating in the AdS space
\be\label{eq:beta}
\beta=\omega \tau,\;\; \alpha =  \pi/2 ,
\ee
and for consistency of the anstaz it turns out that we cannot have
the membrane rotating in the internal sphere at the same time, so the
configuration will not have any R-charge,
\be
\psi= 0,\;\; \theta = \xi = \pi/2, \;\; \phi= 0 .
\ee
Finally, we wrap the membrane along a $U(1)$ in the sphere
\be\label{eq:lambda}
{\tilde{\psi}}= \lambda \delta,\;\; \tilde{\theta} = \pi/2,\;\;  
{\tilde{\phi}}= 0 .
\ee
Note that in (\ref{metrica}), the size of the $M_7$ space does not
change with the AdS radial direction $\rho$ and therefore the wrapped
direction remains stabilised at infinity. This will be the main difference with
the type II solutions below.

We can check that two of the constraints (\ref{eq:constraints}) are satisfied
\beq
G_{\mu\nu}\partial_\sigma X^\mu \partial_\tau X^\nu= 
G_{\mu\nu}\partial_\delta X^\mu \partial_\tau X^\nu= 0 ,
\eeq
whilst the remaining constraint
\beq
\frac{1}{L^2} G_{\mu\nu}\partial_\tau X^\mu \partial_\tau X^\nu= 
(G_{\mu\nu}\partial_\sigma X^\mu \partial_\delta X^\nu)^2 - 
(G_{\mu\nu}\partial_\sigma X^\mu \partial_\sigma X^\nu)
( G_{\mu\nu}\partial_\delta X^\mu 
\partial_\delta X^\nu) ,
\label{constraint}
\eeq
gives the following relation, upon choosing $L = 1/\lambda$,
\beq
\left( \frac{d\rho}{d\sigma} \right)^2= \frac{1}{l_{11}^2 B^2} 
\left[ \kappa^2 \cosh^2\rho - \omega^2 \sinh^2\rho \right] .
\eeq
We may now compute the action by substituting into the formulae of section 2,
\beq
I= - P\int d\tau 
\int_0^{\rho_0} d\rho  \sqrt{\kappa^2 \cosh^2\rho - 
\omega^2 \sinh^2\rho} ,
\label{action1}
\eeq
where $P= \frac{16 \pi |B|}{(2 \pi)^2}$ is a 
normalization 
factor and  $\rho_0$ is the 
turning point given by
\beq                               
\left. \frac{d\rho}{d\sigma}\right|_{\r_0}=0 \Leftrightarrow
\tanh\rho_0= \frac{\kappa}{\omega} .
\label{eq:turning}
\eeq
Note that the term in the action associated to the 
three form vanishes. There is a factor of four in the normalisation
because of the periodicity of the integrand and the fact that the
membrane doubles back on itself.
Write the integrals defining the conserved energy and
spin by differentiating the Lagrangian
\beq
E= -\frac{\delta I}{\delta \kappa}= P\kappa \int_0^{\rho_0} d\rho \frac{\cosh^2\rho}{\sqrt{\kappa^2 
\cosh^2\rho - \omega^2 \sinh^2\rho} },
\label{e1}
\eeq
\beq
S= \frac{\delta I}{\delta \omega}= P\omega \int_0^{\rho_0} d\rho 
\frac{\sinh^2\rho}{\sqrt{\kappa^2 
\cosh^2\rho - \omega^2 \sinh^2\rho} },
\eeq
Now one needs to do the integrals. Fortunately, 
these integrals are exactly the ones considered for rotating
strings \cite{Gubser:2002tv}, and therefore we
may just read off the results from these papers. What one is interested
in is the relationship between the spin and energy for large
and small energy.
In particular, for long membranes we will get the result
\be
E - S = P \ln\frac{S}{P} + \cdots .
\ee
To do the integrals one uses the endpoint constraint
(\ref{eq:turning}) and also one sometimes needs to use the normalisation condition
\be
2\p = \int d\s = \int_0^{r_0} d\rho \frac{d \s}{d \rho} .
\label{j1}
\ee
It is perhaps not surprising that a membrane wrapped on a cycle of constant size has
the same behaviour as a string.

\subsection{Solutions of type II}

We consider now a configuration that is similar to the configuration
of the previous subsection, but with the difference that the wrapped direction is
in the AdS space and not the sphere. That is
\be
\beta=\lambda \delta,\;\; \alpha= \pi/2,
\label{eq:solution2}
\ee
compare this with (\ref{eq:beta}) and (\ref{eq:lambda}).
The rotation must now be in the sphere only, as there are no more directions
in the AdS
\be
{\tilde{\psi}}= \nu \tau, \;\; {\tilde{\theta}}= \xi= \pi/2,\;\; {\tilde{\phi}}= 0.
\ee
The remaining directions are then
\be
t= \kappa \tau,\;\; \rho= \rho(\sigma),\;\; \theta = \pi/2,\;\; \phi =0,\;\; \psi
= 0 .
\ee
The first two constraints are 
satisfied as before, whilst (\ref{constraint}) gives
\beq
\left( \frac{d\rho}{d\sigma} \right)^2 = \frac{1}{l_{11}^2 \sinh^2\rho}\left[ \kappa^2 \cosh^2\rho  - 
B^2\nu^2\right] .
\eeq
The action will be
\beq
I= - P \int d\tau \int_0^{\rho_0} d\rho \sinh\rho \sqrt{\kappa^2 \cosh^2\rho  
-B^2\nu^2} ,
\label{action2}
\eeq
the limits of integration are zero and $\rho_0$ is the solution of the
endpoint constraint, which is now $\cosh^2\rho_0
=\frac{B^2\nu^2}{\kappa^2}$. The normalization 
constant is $P=\frac{8 \pi |B|}{(2 \pi)^2}$ 
and the contribution of 
the $C^{(3)}$ field vanishes as before.

One can now write down the integrals 
defining the energy and R-charge angular momentum, there is no room 
for spin in this case due to the fact that we 
are dealing with $AdS_4$,
\beq
E= -\frac{\delta I}{\delta \kappa}= 
P\kappa \int_0^{\rho_0} d\rho \frac{\sinh\rho\cosh^2\rho}{\sqrt{\kappa^2 
\cosh^2\rho  -
B^2\nu^2} },
\label{e2}\eeq
\beq
J= \frac{\delta I}{\delta \nu}= P B^2 \nu \int_0^{\rho_0} d\rho
\frac{\sinh\rho}{\sqrt{\kappa^2 \cosh^2\rho - B^2\nu^2} }.
\label{j2}\eeq
Again, we now recognise these integrals from previous work, this time
from rotating membranes \cite{Alishahiha:2002sy}. Thus
we may again just read off the energy-R-charge relations. For long
membranes, these are of the form $E = J +...$.

A type II configuration for membranes in $AdS_7\times S^4$ has been
discussed in \cite{Alishahiha:2002sy}. Clearly, one can also write
down a type I configuration in $AdS_7\times S^4$ and obtain a
logarithmic $E-S$ in that background.

\subsection{The case of $AdS_4 \times Q^{1,1,1}$}

We consider now the case where the internal manifold 
$M_7$ of equation (\ref{metrica}) is $Q^{1,1,1}$. This manifold is a 
$U(1)$ fibration over $S^2 \times S^2\times S^2$. The interest of this
configuration is that it provides 
an M theory dual to a three dimensional ${\mathcal{N}}=2$ 
conformal field theory. This is an interesting field theory, that can be 
thought of as describing low energy excitations  living on  M2 branes, 
that are
placed on the tip of an eight dimensional cone with special holonomy. The 
theory is described in terms of fields $A_i, B_i, C_i$ with $i=1,2$
and with given transformation properties under the colour and flavour groups. Gauge 
invariant operators are of the form $X= ABC$ and can be put in 
correspondence with supergravity modes in $AdS_4$. Besides, baryonic 
operators can be constructed.
This theory was well studied in various 
papers \cite{Fabbri:1999hw}, \cite{Herzog:2000rz}, \cite{Gursoy:2002tx}. 

The eleven dimensional configuration reads
\ba
\frac{1}{l_{11}^2} ds_{11}^2 & = & - \cosh^2\rho dt^2 + d\rho^2
\nonumber \\
& + & \sinh^2\rho (d\alpha^2 + \sin^2\alpha d\beta^2) +  
\frac{1}{8} 
(d\Omega_2(\theta_1,\phi_1)+ d\Omega_2(\theta_2,\phi_2)+d\Omega_2(\theta_3,\phi_3) ) \nonumber \\
& + & \frac{1}{16}(d\psi + \cos\theta_1 d\phi_1 + \cos\theta_2 d\phi_2 + \cos\theta_3 d\phi_3 )^2 ,
\label{metricaq111}
\ea
\beq
H = k \cosh\rho \sinh^2\rho \sin\alpha dt \wedge d\rho \wedge d\alpha
\wedge d\beta,\;\;\; C= - \frac{k 
\sin\alpha}{3} \sinh^3\rho dt \wedge d\alpha\wedge d\beta
\label{gaugefieldq111}
\eeq
Here again, $k$ is a constant determined by the equations of motion.

We can again consider two types of solutions. We will be 
brief in this case, since the calculations results are not 
very different from those of the previous subsections. Indeed the main
point here is that the existence of two types of energy-spin relations,
one with logarithms and one without, is independent of the internal manifold,
so long as it has a $U(1)$ isometry around which we can wrap the membrane.

\begin{itemize}
\item Type I solutions
\end{itemize}

\noindent In this case the configuration reads reads
\beq
t= \kappa \tau,\;\; \rho= \rho(\sigma), \;\; \beta=\omega \tau, \;\; 
\alpha =  {\theta_i}=\pi/2, \;\;
{{\phi_i}}= \nu_i \tau,\;\; \psi= \lambda\delta .  
\label{solution1q111}
\eeq
The constraint reads
\beq
\left(\frac{d\rho}{d\sigma}\right)^2= 
\frac{16}{l_{11}^2}\left(-\omega^2 \sinh^2\rho + \kappa^2\cosh^2\rho - 
\frac{1}{8} (\nu_1^2 + \nu_2^2 + \nu_3^2)\right) ,
\eeq
there is a turning point where $d\rho/d\sigma=0$
and the action reads
\beq
I = -P \int d\tau  \int_0^{\rho_0} d\rho \frac{1}{4}  
\sqrt{\kappa^2 \cosh^2\rho - \omega^2 \sinh^2\rho -
\frac{1}{8} (\nu_1^2 + \nu_2^2 + \nu_3^2)} .
\label{action1q111}
\eeq
In this case, the normalization factor is $P= \frac{16\pi}{(2\pi)^2}$. As in 
the previous sections, the contribution of the $C^{(3)}$ field vanishes.
We can compute the energy, spin and 
R-charge angular momentum, and the results are essentially identical to those coming
from equations (\ref{e1})-(\ref{j1}). In particular, there is a logarithmic $E-S$ relation.

\begin{itemize}
\item Type II solutions
\end{itemize}

\noindent In this case the solution will read
\beq
t= \kappa \tau,\;\; \rho= \rho(\sigma),\;\; \beta=\lambda \delta, \;\; 
\alpha = {\theta_i}=\pi/2,\;\;, \psi = 0,\;\;
{{\phi_i}}= \nu_i \tau .  
\label{solution2q111}
\eeq
The constraint will give a turning point $\rho_0$, when $d\rho/d\sigma=0$
\beq
\left(\frac{d\rho}{d\sigma} \right)^2 l_{11}^2 \sinh^2\rho = \kappa^2 \cosh^2\rho-\frac{1}{8} 
(\nu_1^2 + \nu_2^2 + \nu_3^2) ,
\eeq
and the action will be
\beq
I= P \int d\tau \int_0^{\rho_0} d\rho \sinh\rho 
\sqrt{\kappa^2 \cosh^2\rho  -
\frac{1}{8} (\nu_1^2 + \nu_2^2 + \nu_3^2)} .
\label{action2q111}
\eeq
The normalization factor will be $P= \frac{8\pi}{(2\pi)^2}$.
This time, the  results will be essentially the same as those coming from
the previous type II configuration, of equations (\ref{e2})-(\ref{j2}).

It is not difficult to see from (\ref{metricasquashed}) that one may
obtain the same results using the squashed seven sphere, as we have
the same cycles on which to wrap the membrane and rotate.
Thus we obtain membrane configurations dual to operators of an ${\mathcal{N}}=1$
theory in three dimensions. 

We can also consider
the case of three dimensional ${\mathcal{N}}=3$ conformal field theories. 
These 
theories are dual to geometries of the form $AdS_4 \times N^{0,1,0}$, 
where the manifold $N^{0,1,0}$ has metric
\beq
ds^2_7 = d\xi^2 + \frac{1}{4} \sin^2\xi (\sigma_1^2 + \sigma_2^2 + \cos^2\xi 
\sigma_3^2) + \frac{1}{2}(\omega^i - A^i)^2
\eeq
with $A^{1,(2)} = \cos\xi \sigma_{1,(2)}$, $2 A^{3}=  (1 + 
\cos^2\xi)\sigma_3$, and $\omega^i, \sigma_i$ are left-invariant forms in 
the different $SU(2)$s. This type of field theory is interesting because  
it has the 
same field content as ${\mathcal{N}}=4$ theories, but there are fermionic interactions 
that only preserve ${\mathcal{N}}=3$. 
Following the steps above, one can find type I and type II solutions for 
these metrics. Everything will work as before, with different 
numerical coefficients.

It should be clear by now that all that is needed to obtain a
logarithmic configuration is a stabilised circle to wrap the membrane
on. As we will see below in the section on $G_2$ manifolds, this does
seems work in more general situations than AdS product spacetimes.

\subsection{Membranes moving in warped $AdS_5\times M_6$ spaces}

We now consider membranes moving in a geometry that is dual 
to an ${\mathcal{N}}=2$ supersymmetric conformal field theory in four dimensions,
as opposed to the three dimensions of the previous cases.
The eleven dimensional configuration was written in 
\cite{Maldacena:2000mw}
and represents M5 branes wrapping a hyperbolic two-manifold. The 
geometry has the form of a warped product of five dimensional AdS space times a six 
dimensional manifold. This should be thought of as M5 branes wrapping some 
compact (hyperbolic) cycle
inside a Calabi-Yau two-fold.

The metric looks, schematically 
(for a detailed discussion see \cite{DiVecchia:2001uc}), as follows
\ba
\frac{1}{l_{11}^2} ds^2 & = & \Delta(\theta)^{1/3} (R_a^2 ds^2_{AdS_5}
+ R_b^2 (d\tilde{\theta}^2 + 
\sinh^2\tilde{\theta} d\tilde{\phi}^2) +R_c^2 d\theta^2 
+\nonumber\\
& &R_c^2\frac{\Delta(\theta)^{-2/3}}{4}( 
\cos^2\theta (d\alpha^2 + \sin^2\alpha d\beta^2) + 2\sin^2\theta (d\psi - 
\cosh\tilde{\theta} d\tilde{\phi})^2  ) ,
\ea
where $AdS_5$ is written in the coordinates $(\rho, t, \xi_1, \xi_2, 
\phi)$ as usual, $\Delta(\theta) = 1+\cos^2\theta$ and the  $R_i$ are constants.
The $C^{(3)}$ field has the schematic form
\beq
C= f_1(\alpha, \theta, \tilde{\theta}) d\tilde{\theta} \wedge 
d\tilde{\phi}\wedge d\beta + f_2(\alpha, \theta) d\theta \wedge d\beta 
\wedge (d\psi - \cosh\tilde{\theta} d\tilde{\phi}) .
\eeq
We can consider a solution of the form
\beq
\rho= \rho(\sigma), \;\;t=\kappa \tau, \;\;\phi= \omega\tau, \;\; 
\psi=\lambda \delta,\;\; \theta=\frac{\pi}{2} .
\eeq
The remaining angles take values of $0$ or $\pi/2$ as in previous
configurations.
Note that $\theta=\pi/2$ is necessary to solve the equations of
motion. There is no R-charge in this configuration. Warped solutions are
discussed in more detail in Appendix A 
below. The key point about this configuration is that the warping
factor is unimportant because we fix a value of $\theta$ so that
it just becomes an overall number. The configuration is
of type I because the wrapped direction, $\psi=\lambda \delta$,
is in the $M_6$, therefore we expect to get relations of the form
$E- S \sim \ln S$, and indeed this is what one finds upon doing
the calculations. The integrals that emerge are, up to
numerical coefficients, the same as the type I configurations we
studied above.

\subsection{Membranes moving near an AdS black hole}

For completeness, we briefly
consider now the case of membranes orbiting in an eleven dimensional geometry given 
by
\beq
\frac{1}{l_{11}^2} ds^2_{11}= - f(r)dt^2 + \frac{dr^2}{f(r)} + r^2 (d\alpha^2 + \sin^2\alpha 
d\beta^2)+ B^2 ds^2_{S^7} .
\label{metrihole}
\eeq
With the function
\be
f(r)= 1 + r^2/R^2 - M^2/r ,
\ee
and $ B=R$, the previous  metric  
is a black hole in $AdS_4$.     
A very nice physical description of the AdS/CFT correspondence for strings orbiting
about black holes was given in \cite{Armoni:2002xp}. We shall limit ourselves to discussing the
membrane configurations and energy spin relations.
As previously, we will consider two types of configuration, the type I
\beq
t= \kappa \tau,\;\; \rho= \rho(\sigma), \;\; \beta= \omega \tau, \;\; 
\alpha=  {\theta_i}= {\tilde{\theta}_i}=\pi/2, \;\;
{\tilde{\psi }}= \lambda \delta,\;\;\xi=\pi/4 ,
\label{solution1black}
\eeq
and the type II
\beq
t= \kappa \tau,\;\; \rho= \rho(\sigma), \;\; \beta=\lambda \delta, \;\; 
\alpha = {\theta_i} = {\tilde{\theta}_i} =\pi/2, \;\;
{\tilde{\psi }}= \nu \tau .  
\label{solutionblack2}
\eeq
We can construct the expression for the membrane constraint in the first case,
\beq
2 f(r) (-\omega^2 r^2 + \kappa^2 f(r))= l_{11}^2 B^2\left(\frac{dr}{d\sigma}\right)^2 ,
\eeq
and for the type II solutions
\beq
f(r) (-\nu^2 B^2 +2 \kappa^2 f(r))= 2 r^2 l_{11}^2 \left(\frac{dr}{d\sigma}\right)^2 .
\eeq
Upon requiring $dr/d\sigma$ to vanish at the endpoints,
we will obtain two different values of $r_{min}, r_{max}$, that is, 
the integration limits in the action, when we change variables from $\sigma$ 
to the radial coordinate $r$. This is physically the fact that the membrane is entirely
outside the event horizon and is therefore orbiting rather than rotating.

The expression for the action, 
energy and spin, in the type I case is,
\beq
I= P \int d\tau 
\int_{r_{min}}^{r_{max}} 
\sqrt{\kappa^2 f(r) - \omega^2 r^2}\frac{dr}{\sqrt{f(r)}},
\eeq
\beq
E= \kappa P 
\int_{r_{min}}^{r_{max}} 
\frac{ \sqrt{f(r)}}{\sqrt{\kappa^2 f(r) - \omega^2 r^2}}dr,
\eeq
\beq
S= \omega P \int_{r_{min}}^{r_{max}} \frac{r^2}{\sqrt{\kappa^2 f^2(r) - 
\omega^2 r^2 f(r)}}dr,
\eeq
while for the type II configurations we will have for the action, energy 
and R-symmetry angular momentum
\beq
I= \frac{P}{2} \int d\tau   
\int_{r_{min}}^{r_{max}} 
\sqrt{\kappa^2 f(r) - \nu^2 B^2/2}\frac{r}{\sqrt{f(r)}}dr ,
\eeq
\beq
E= \frac{\kappa P}{2} \int_{r_{min}}^{r_{max}} \frac{ r  
\sqrt{f(r)}}{\sqrt{\kappa^2 f(r) - \nu^2 B^2/2}}dr ,
\eeq
\beq
J= \frac{\nu P B^2}{4} \int_{r_{min}}^{r_{max}} \frac{r}{\sqrt{\kappa^2 
f^2(r) - \nu^2 B^2 f(r)/2}}dr .
\eeq
Let us study the explicit expressions for the energy and the spin of type 
I configurations. It is convenient to make a choice of parameters 
$M=R=\kappa=1$. The results will remain true at least for a small
interval of value for $M$ around $M=1$.
We can see that for values of the parameter
$\omega$ close to one, corresponding to long membranes, the functions
inside the square roots are positive 
in the interval $r_h, r_+$, where $r_h$ is the root of $r^3 + r 
-1$ and the roots of $-(\omega^2-1)r^3 + r -1$ are $r_-, r_+$ both of 
them  positive and bigger than $r_h$ and $r_*$, a third negative root.
The integrals read,
\beq
E\sim \int_{r_-}^{r_+} dr \sqrt{\frac{r^3 + r -1}{-(\omega^2-1)r^3 + r 
-1}} ,
\eeq
\beq
S\sim  \int_{r_-}^{r_+} dr  \frac{r^3}{\sqrt{[r^3 + r -1] 
[-(\omega^2-1)r^3 + r -1]}} .
\eeq
Now, we want to study the approximate expressions for these integrals in 
the cases of long membranes, that is membranes that are extended in the 
interval $(r_- , r_+)$.
Evaluating the approximate expressions for the integrals in the
case of long membranes, we get a relation of the form $E- k S\sim
S^3$.

\section{Rotating membranes on $G_2$ manifolds}

\subsection{The duality with ALC $G_2$ metrics}

Partially motivated by the developments described in the introduction,
there has been significant recent progress in constructing new
cohomogeneity-one manifolds with $G_2$ holonomy
\cite{Brandhuber:2001yi,Cvetic:2001ya,Cvetic:2001zx,Cvetic:2001sr,
Cvetic:2001bw,Cvetic:2001kp,Cvetic:2001ih,
Brandhuber:2001kq}, generalising the manifolds that have been known
for some time \cite{bs,Gibbons:er}.

When the M-theory flop was discussed in \cite{Atiyah:2000zz}, the only known $G_2$ metric
with the necessary symmetries to describe wrapped D6 branes in type IIA was
asymptotically a cone over $S^3\times S^3$ \cite{bs,Gibbons:er}. The
essential point is that one $S^3$ collapses at the origin
whilst another does not. \footnote{Note that the collapsing and non-collapsing $S^3$s need
not coincide with the two $S^3$s in terms of which the metric is written.}
Thus depending on which $S^3$ the M-theory
cycle is contained in, one gets either a IIA reduction that is singular
at the origin - branes - or
a non-singular reduction - no branes. However, in these metrics the
dilaton diverges at infinity after reduction so they are
unsatisfactory IIA backgrounds. The authors of \cite{Atiyah:2000zz}
thus postulated the existence of two new types of $G_2$ holonomy
metric to fix this problem. These metrics should not be Asymptotically
Conical (AC), but Asymptotically Locally Conical (ALC), that is to say
that at infinity there should be a circle with a stabilised
radius. This circle will be the M-theory circle and the corresponding
IIA dilaton will be well-behaved. The two metrics would correspond to when
the stabilised $U(1)$ is contained within the collapsing $S^3$ or the
non-collapsing $S^3$, corresponding to good D6-brane or good non-D6-brane
solutions, respectively.

This picture was essentially realised with the discovery of explicit ALC
$G_2$ metrics. $G_2$ metrics reducing to D6-branes wrapping the
deformed conifold were discussed in
\cite{Brandhuber:2001yi,Cvetic:2001bw,Cvetic:2001zx}, these are called
the $\B_7$ family of metrics. Metrics reducing to the small
resolution of the conifold with fluxes were discussed in
\cite{Cvetic:2001kp,Brandhuber:2001kq,Cvetic:2001ih}, these are called
the ${\Bbb D}_7$ family. Transformations of these metrics under the
broken triality symmetry were discussed in \cite{Brandhuber:2001kq},
this does not change the radial behaviour or the symmetries.

The situation is not quite as anticipated by \cite{Atiyah:2000zz}. All
the known $G_2$ metrics are constructed out of left-invariant one-forms on
the two $S^3=SU(2)$s, $\{\S_i,\s_i\}_{i=1}^3$, see e.g. (\ref{eq:b7metric})
below. In the AC case, there is an $SU(2)_L^\S\times SU(2)_L^\s\times
SU(2)_R^D$ isometry group, where the $SU(2)_L^\S\times SU(2)_L^\s$ part is manifest
and corresponds to left multiplication on the spheres. The remaining
$SU(2)_R^D$ is a diagonally acting right multiplication. Note that
right multiplication acts on the left-invariant forms as an adjoint
action. The ALC metrics have a reduced isometry group $SU(2)\times
SU(2) \times U(1)$, where an $SU(2)$ has been broken to
$U(1)$ by the stabilised cycle. It was suggested in
\cite{Atiyah:2000zz} that the $SU(2)$ that should be broken to $U(1)$
would be $SU(2)_L^\S$ on one side of the flop and $SU(2)_L^\s$ on the
other side. This fits nicely with symmetry of their
discussion. However, in order to realise this, one would need to
construct $G_2$ metrics that were not written in terms of left
invariant forms, as these automatically imply $SU(2)_L^\S\times
SU(2)_L^\s$ invariance. It is not clear how one would go about doing
this. Instead, the solutions of the $\B_7$ and ${\Bbb D}_7$
families have the $SU(2)_R^D$ broken to $U(1)$.
This is compatible with writing the metric in terms of the left invariant one forms.

There is a ${\mathbb{Z}}_N$ quotient of the metric that is responsible for
the $N$ D-branes or the $N$ units of flux upon reduction to IIA.
The ${\mathbb{Z}}_N$ always acts within the diagonal $U(1)$ which is
furthermore the M-theory circle. The action is singular at the origin
in the $\B_7$ case, but not for ${\Bbb D}_7$.

It is important to observe that both the $\B_7$ and ${\Bbb D}_7$ families
are two parameter families and one can go from one to the
other \cite{Cvetic:2001ih}, via the singular AC metric. Besides the scale
parameter which measures the 
distance from the singular conical point, there is another parameter
which measures the distance from 
the AC metric. The AC metric is contained in both families. Consider now the
running of the coupling constant, described in the 
first section. Starting in a $\B_7$ metric in the UV, the
flow must involve not only one of the $S^3$s shrinking
- change of scale parameter - but also a flow towards the AC metric.
This allows the flow to move to the
${\Bbb D}_7$ family via the AC metric as well as expanding a different
$S^3$ in the IR. Thus the flow must move nontrivially in a two
dimensional parameter space. A priori, it is not obvious why starting
from an M-theory geometry that has a well behaved dilaton in the IIA
reduction in the UV ($\B_7$ family), one should generically end up with an M-theory
geometry that also admits a good IIA reduction in the IR (${\Bbb D}_7$
family). But the desired flow should exist, which is enough to establish
the IIA duality from M-theory with a well-behaved dilaton. Assuming
that the quantum smoothing of the process continues to occur as it did
in the AC case \cite{Atiyah:2001qf}.

We will consider membranes rotating on all of the geometries discussed
in this subsection. The ${\Bbb D}_7$ family are, strictly speaking,
the gravity duals that describe the field theory in the IR. The precise
role of the $\B_7$ metrics in the duality is unclear, although it
could well be related to the lack of brane-bulk decoupling
discussed in the introduction.

\subsection{Membranes on the Asymptotically Conical metric and the $\B_7$
family}

\subsubsection{Metric formulae}

The background is pure geometry, the three-form C-field is zero.
The eleven dimensional metric is of the form
\be
\frac{1}{l_{11}^2} ds^2_{11} = -dt^2 + dx^2 + dy^2 + dz^2 + ds^2_7 ,
\ee
where the $G_2$ metrics are
\bea\label{eq:b7metric}
ds^2_7 = dr^2 + a(r)^2\left[(\S_1 - \s_1)^2
+(\S_2 - \s_2)^2\right] + d(r)^2(\S_3 -\s_3)^2 \nn \\
+ b(r)^2 \left[(\S_1 + \s_1)^2 +(\S_2 + \s_2)^2 \right]
+ c(r)^2(\S_3 + \s_3)^2 ,
\eea
where $\S_i,\s_i$ are left-invariant one-forms on $SU(2)$
\bea\label{eq:1forms}
\s_1 & = & \cos\psi_1 d\q_1+\sin\psi_1\sin\q_1 d\phi_1 , \nn \\
\s_2 & = & -\sin\psi_1 d\q_1+\cos\psi_1\sin\q_1 d\phi_1 , \nn \\
\s_3 & = & d\psi_1 + \cos\q_1 d\phi_1 ,
\eea
where $0\leq\q_1\leq\p$, $0\leq\phi_1\leq 2\p$, $0\leq\psi_1\leq 4\p$,
at least before including the ${\mathbb{Z}}_N$ quotient.
The definitions for $\S_i$ are analogous but with
$(\q_1,\phi_1,\psi_1) \to (\q_2,\phi_2,\psi_2)$.
These metrics are locally asymptotic to cones over $S^3\times S^3$. There is
a finite size $S^3$ bolt at the origin.
There is a two parameter family of such
$G_2$ metrics, called $\B_7$ in the classification of \cite{Cvetic:2001kp,Cvetic:2001ih}.
The radial functions satisfy the equations \cite{Brandhuber:2001yi,Cvetic:2001zx}
\bea\label{eq:equations}
\dot{a} =
\frac{1}{4}\left[-\frac{a^2}{bd}+\frac{d}{b}+\frac{b}{d}+\frac{c}{a}\right]
& , & \;\;
\dot{b} = \frac{1}{4}\left[-\frac{b^2}{ad}+\frac{d}{a}+\frac{a}{d}-\frac{c}{b}\right], \nn \\
\dot{d} = \frac{1}{2}\left[-\frac{d^2}{ab}+\frac{a}{b}+\frac{b}{a}
\right] & , & \;\;
\dot{c} = \frac{1}{4}\left[\frac{c^2}{b^2} -\frac{c^2}{a^2}\right].
\eea
Two exact solutions are known. One is the asymptotically conical (AC)
solution of \cite{bs,Gibbons:er}, which has $SU(2)^3\times \Z_2$
symmetry. The other is only Asymptotically Locally Conical (ALC), with
a stabilised $U(1)$ at infinity
\cite{Brandhuber:2001yi,Cvetic:2001bw}, which has $SU(2)^2\times
U(1)\times \Z_2$ symmetry. 
The remaining metrics in this family are only known
numerically. Fortunately, we only require the asymptotics at the origin
and at infinity, which are easily calculated from
(\ref{eq:equations}). As $r \to 0$ we have
\bea\label{eq:bshort}
a(r) & = &  R_0 + \frac{r^2}{16 R_0} - \frac{(7+64q_0) r^4}{2560 R_0^3}  + \cdots, \nn \\
b(r) & = &  \frac{r}{4} + \frac{q_0 r^3}{R_0^2} - \frac{(-1+98304 q_0^2+1344q_0)r^5}{10240 R^4_0} + \cdots , \nn \\
c(r) & = &  \frac{r}{4} - \frac{(1+128q_0) r^3}{64 R_0^2} +
\frac{(216q_0+1+16896 q_0^2)r^5}{640 R_0^4} + \cdots , \nn \\
d(r) & = & R_0 + \frac{r^2}{16 R_0} + \frac{(-3+64 q_0)r^4}{1280 R_0^3} + \cdots ,
\eea
where $q_0$ and $R_0$ are constants. Note that $b(r)$ and $c(r)$ collapse,
whilst $a(r)$ and $d(r)$ do not.
As $r\to\infty$ we have
\bea\label{eq:blong}
a(r) & = & \frac{r}{2\sqrt{3}} + \frac{R_1}{2} (2 q_1 + \sqrt{3}) +
\frac{3\sqrt{3}R_1^2}{4r} +
\cdots, \nn \\
b(r) & = & \frac{r}{2\sqrt{3}} + q_1 R_1 + \frac{3\sqrt{3}R_1^2}{4r} + \cdots  , \nn \\
c(r) & = & R_1 - \frac{9 R^3_1}{r^2} + \frac{(27+36\sqrt{3} q_1) R_1^4}{r^3} + \cdots, \nn \\
d(r) & = & \frac{r}{3} + \frac{R_1}{2\sqrt{3}} (4q_1+\sqrt{3}) +
\frac{3 R_1^2}{r} + \cdots,
\eea
where $q_1$ and $R_1$ are constants that will be functions of
of $q_0$ and $R_0$. Note that $c(r)$ is stabilised whilst the others
diverge linearly.
The expressions are needed to second order because we will be
interested in the subleading terms of various integrals.

\subsubsection{Commuting $U(1)$ isometries and membrane configurations}

The metrics (\ref{eq:b7metric}) have three commuting $U(1)$
isometries. Using the Euler coordinates (\ref{eq:1forms}), these can
canonically be taken to be generated by $\del_{\phi_1} \subset SU(2)_L^{\s}$, $\del_{\phi_2}
\subset SU(2)_L^{\S}$ and $\del_{\psi_1}+\del_{\psi_2} \subset
SU(2)_R^D$.

The existence of three commuting $U(1)$ isometries is very useful for considering
rotating membranes. By placing the directions of rotation and wrapping
along these $U(1)$s, most of the equations of motion are trivially
satisfied as a statement of conserved charges. The remaining equation
of motion for the radial direction then follows from a first order
gauge fixing constraint, as discussed above.

However, the canonical $U(1)$s are not the most sensible for our
purposes. Consider the redefinitions
\bea
\psi_3 = \psi_1 + \psi_2, & \;\; & \psi_4 = \psi_1 - \psi_2, \nn \\
\phi_3 = \phi_1 + \phi_2, & \;\; & \phi_4 = \phi_1 - \phi_2 .
\eea
Note that $\psi_3,\psi_4$ now have a range of $8\p$ whilst
$\phi_3,\phi_4$ have a range of $4\p$.
Three commuting isometries now are $\del_{\phi_3},
\del_{\phi_4},\del_{\psi_3}$. As we shall see shortly, the first two
$U(1)$s will now be contained in $S^3$s that do collapse and do not collapse,
respectively, at the origin. In the IIA brane picture, the $S^3$ that
does collapse is surrounding the brane whilst the $S^3$ that does not
collapse is inside the brane. In order for the dual field
theory to be four dimensional, one must consider energies such that
the finite $S^3$ is not probed.
The charge generated by rotations along
$\del_{\phi_3}$, inside the brane, will be 
denoted $K_1$, whilst the charge generated by rotations along
$\del_{\phi_4}$, outside the brane, will be denoted by
$K_2$. In the $\B_7$ family, the $U(1)$ generated by
$\del_{\psi_3}$, the circle that is stabilised at infinity, is contained within the
collapsing $S^3$ at the origin. Call this charge $K_3$. Note that the isometries
transverse to the membrane do not have the interpretation of R-charge
because the theory is ${\mathcal{N}}=1$.
We cannot have all three charges at once, as we need to use
one of the isometries to wrap the membrane. This last point is
necessary for the wrapping direction to drop out of the action integral.

The configuration of the membrane is taken to be trivial in the
remaining directions
\be
x = y = z = 0, \;\; \q_1 = \q_2 = \p/2, \;\; \psi_4 = 0 .
\ee
One can also extend the string in the $xyz$ plane and indeed such
configuration will be considered in a later section.
There are then three possible configurations for the nontrivial
directions, shown in Table 1.

\begin{table}[h]
{\bf Table 1: } Three rotating membrane configurations on the $\B_7$ metrics \\
  \begin{tabular}{|c|c|c|c|} \hline
Target space coordinate & Configuration I$_\B$ & Configuration II$_\B$ &
  Configuration III$_\B$ \\ \hline \hline
$\phi_3 =$ & $\w \t$ & $\l \d$    & $\w \t$ \\ \hline
$\phi_4 =$ & $\nu_2 \t$    & $\nu_2 \t$    & $\l \d$ \\ \hline
$\psi_3 =$ & $\l \d$    & $\nu_3 \t$ & $\nu_3 \t$ \\ \hline
$t =$ & \multicolumn{3}{|c|}{$\k \t$} \\ \hline
$r =$ &  \multicolumn{3}{|c|}{$r(\s)$} \\ \hline
  \end{tabular}
\end{table}

If the stabilised circle generated by $\del_{\psi_3}$ is considered as
the M-theory circle, then configurations II$_\B$ and III$_\B$ reduce
to rotating D2-branes or a D0-D2 state, depending on whether there is a rotation
along the M-theory circle or not, whilst the I$_\B$ configuration reduces to a
rotating fundamental string.

At this point, we need to take into account the ${\mathbb{Z}}_N$
quotient of the $G_2$ manifold that was mentioned above. The effect of
this quotient is to send
\be
\psi_3 \to \frac{\psi_3}{N} .
\ee
The target space metric that is seen by the membrane is thus
\be
\frac{1}{l_{11}^2} ds^2_{M2} = -dt^2 + dr^2 + \frac{c(r)^2}{N^2} d\psi_3^2 + b(r)^2
d\phi_3^2 + a(r)^2 d\phi_4^2 .
\ee
It is easily checked that the $\g_{0\a}$ constraints in
(\ref{eq:constraints}) are satisfied. The remaining constraint,
choosing the free constant $L=1/\l$, then implies that
\be
\left(\frac{d r}{d \s}\right)^2 =
\left\{
\begin{array}{r}
\frac{\k^2 - b(r)^2 \w^2 - a(r)^2 \nu_2^2}{c(r)^2 l_{11}^2/N^2} \quad\quad (\mbox{Case I$_\B$}) \\
\frac{\k^2 - a(r)^2 \nu_2^2 - \nu_3^2 c(r)^2 / N^2}{b(r)^2 l_{11}^2}  \quad\quad (\mbox{Case II$_\B$}) \\
\frac{\k^2 - b(r)^2 \w^2 - c(r)^2 \nu^2_2/N^2}
{a(r)^2 l_{11}^2} \quad\quad (\mbox{Case III$_\B$})
\end{array}
\right.
\ee
A further constraint must be imposed, this is the condition that the
membrane doubles back on itself at some radius
\be\label{eq:endpoint}
\left.\frac{d r}{d \s}\right|_{r_0} = 0 .
\ee
This condition gives a relationship between
$(r_0,\k,\w,\nu_2,\nu_3)$. We will use this relationship to eliminate
$\kappa$ below.

Also, one needs to impose a normalisation condition
\be\label{eq:normalisation}
2\p = \int d\s = \int_0^{r_0} dr \frac{d \s}{d r} .
\ee
This gives an integral constraint between $(r_0,\k,\w,\nu_2,\nu_3)$.

Other configurations are possible, in which the rotating
directions or the wrapped direction is some linear combination of the
$U(1)$s. However, these configurations will not generically satisfy the
constraints, because the induced metric will have cross terms. Another
possibility is to take different $U(1)$ subgroups of the original
$SU(2)$s. The present choices would appear to be the most natural and we
will not consider other subgroups here.

Before moving on, one must check the consistency of the ans\"atze
described here. Checking the equations of motion, we find that indeed
all are consistent.

\subsubsection{Energy and other conserved charges}

The following conserved charges are naturally associated with the configuration
\bea
E & = & -\frac{\delta I}{\delta \kappa} , \quad\quad K_1 = \frac{\delta I}{\delta \omega} , \nn \\
K_2 & = & \frac{\delta I}{\delta \nu_2} , \quad\quad K_3 = \frac{\delta I}{\delta \nu_3} , 
\eea
where $I$ is the action (\ref{eq:M2action}). The $\k$ derivative is
taken at fixed $(r_0,\w,\nu_2,\nu_3)$, and similarly for the other
derivatives.

Let us do this in the three cases.
Note that in passing from an integral over $\s$ to an integral over
$r$ we multiply by four because of the periodicity of the integrand
and the fact that the membrane doubles back on itself. We use the
constraint (\ref{eq:endpoint}) to eliminate $\k$. The different
numerical factor in the different cases is due to the different ranges
of the angle about which the membrane is wrapped.

\begin{itemize}
\item Case I$_\B$

\bea
I & = & \frac{-32 \p}{N (2\p)^2}\int d\t\int_0^{r_0} dr |c(r)| \sqrt{\k^2 - b(r)^2
\w^2 - a(r)^2 \nu_2^2} , \\
E & = & \frac{32 \p}{N (2\p)^2}\int_0^{r_0} dr
\frac{|c(r)| \sqrt{\w^2 b(r_0)^2 + \nu_2^2
a(r_0)^2}}{\sqrt{\w^2\left[b(r_0)^2-b(r)^2\right] + \nu_2^2
\left[a(r_0)^2-a(r)^2\right]}} , \\
K_1 & = & \frac{32 \p}{N (2\p)^2}\int_0^{r_0} dr
\frac{\w |c(r)| b(r)^2}{\sqrt{\w^2\left[b(r_0)^2-b(r)^2\right] + \nu_2^2
\left[a(r_0)^2-a(r)^2\right]}}, \\
K_2 & = & \frac{32\p}{N (2\p)^2}\int_0^{r_0} dr
\frac{\nu_2 |c(r)| a(r)^2}{\sqrt{\w^2\left[b(r_0)^2-b(r)^2\right] + \nu_2^2
\left[a(r_0)^2-a(r)^2\right]}}, \\
K_3 & = & 0 .
\eea

\item Case II$_\B$

\bea
I & = & \frac{-16\p}{(2\p)^2}\int d\t\int_0^{r_0} dr
|b(r)| \sqrt{\k^2 - a(r)^2 \nu_2^2 - \nu_3^2 c(r)^2 /N^2} , \\
E & = & \frac{16\p}{(2\p)^2}\int_0^{r_0} dr
\frac{|b(r)| \sqrt{\nu_2^2 a(r_0)^2 + \nu_3^2
c(r_0)^2 / N^2}}{\sqrt{\nu_2^2\left[a(r_0)^2-a(r)^2\right] + \nu_3^2
\left[c(r_0)^2-c(r)^2\right] / N^2}}, \\
K_1 & = & 0, \\
K_2 & = & \frac{16\p}{(2\p)^2}\int_0^{r_0} dr
\frac{\nu_2 |b(r)| a(r)^2}{\sqrt{\nu_2^2\left[a(r_0)^2-a(r)^2\right] + \nu_3^2
\left[c(r_0)^2-c(r)^2\right] / N^2}} , \\
K_3 & = & \frac{16\p}{N^2 (2\p)^2}\int_0^{r_0} dr
\frac{\nu_3 |b(r)| c(r)^2}{\sqrt{\nu_2^2\left[a(r_0)^2-a(r)^2\right] + \nu_3^2
\left[c(r_0)^2-c(r)^2\right] / N^2}} .
\eea

\item Case III$_\B$

\bea
I & = & \frac{-16\p}{(2\p)^2}\int d\t\int_0^{r_0} dr |a(r)|
\sqrt{\k^2 - b(r)^2 \w^2 - \nu_3^2 c(r)^2 /N^2} , \\
E & = & \frac{16\p}{(2\p)^2}\int_0^{r_0} dr
\frac{|a(r)|\sqrt{\w^2 b(r_0)^2 + \nu_3^2
c(r_0)^2 / N^2}}{\sqrt{\w^2\left[b(r_0)^2-b(r)^2\right] + \nu_3^2
\left[c(r_0)^2-c(r)^2\right] / N^2}} , \\
K_1 & = & \frac{16\p}{(2\p)^2}\int_0^{r_0} dr
\frac{\w |a(r)| b(r)^2}{\sqrt{\w^2\left[b(r_0)^2-b(r)^2\right] + \nu_3^2
\left[c(r_0)^2-c(r)^2\right] / N^2}} , \\
K_2 & = & 0 , \\
K_3 & = & \frac{16\p}{N^2 (2\p)^2}\int_0^{r_0} dr
\frac{\nu_3 |a(r)| c(r)^2}{\sqrt{\w^2\left[b(r_0)^2-b(r)^2\right] + \nu_3^2
\left[c(r_0)^2-c(r)^2\right] / N^2}} .
\eea

\end{itemize}
These integrals may be expanded for large and small $r_0$ using the
expansions (\ref{eq:bshort}) and (\ref{eq:blong}). In the various
integrals there are usually two constants, such as $\w$ and $\nu_2$ in
the I$_\B$ case. These are nontrivially related through the
normalisation constraint (\ref{eq:normalisation}). Here we will only
consider the cases where one of the constants is zero, corresponding
to a rotation in only one direction. In these cases we see that the
remaining constant drops out of the integral and the normalisation
constraint does not need to be evaluated.

For short membranes, small $r_0$, one may use the Taylor expansions
about the origin to evaluate the integrals. For long membranes, large
$r_0$, one may only use the expansions about infinity to
evaluate the integral if the integral is diverging with $r_0$ because
in this case the integral is dominated by the contributions at
infinity. One then needs to check that there is not an $r_0$ contribution
from the interior of the integrand. Naively, the integrals for
large $r_0$ are done as follows
\bea\label{eq:exp}
\int_0^{r_0} f(r,r_0) dr & \approx & \int_{\Lambda}^{r_0} f(r,r_0) dr
= r_0 \int_{\Lambda}^{1} f(u r_0,r_0) du \nonumber \\
& \approx & r_0 \int_{\Lambda}^{1} \left[ F_m(u) r_0^m + F_{m-1}(u)
r_0^{m-1} + \cdots \right]  du ,
\eea
where $\Lambda$ is some cutoff and we ignore contributions from this
end of the integral. The final expression represents an expansion of
$f(u r_0,r_0)$ about $r_0 = \infty$. In the final result of this
calculation, we may trust any terms that diverge as $r_0\to\infty$.
One thing that may go wrong is that the leading order
coefficient, $F_m(u)$, in the final equation of (\ref{eq:exp}) integrates to
zero, meaning that there is no $r_0^m$ power term.
In this case one should do the full integral numerically to check
whether the vanishing is a result of power expanding inside the
integral, and see what the leading order coefficient is. Alternatively
one can try to do the integral exactly without expanding the integrand
fully. Doing this is crucial to obtain the logarithmic term in the
next subsection.

Given the resulting expressions for $E$ and the $K$s, one then
eliminates $r_0$ to obtain the results of Table 2. In this table $k$
is used to denote positive numerical factors. Dependence on
$R_0,R_1,N$ is kept explicit. It turns out there is no dependence on
$q_0,q_1$ to the order considered in the table.

\begin{table}[h]
{\bf Table 2: } Energy - Charge relations for membranes on $\B_7$ metrics \\
  \begin{tabular}{|c|c|c|c|} \hline
Configuration & $r_0 \to 0$ (short membranes) & $r_0 \to \infty$ (long
  membranes)\\ \hline \hline
\rule[-5mm]{0mm}{13mm} I$_\B$, $\nu_2 = 0$   & $\displaystyle E = k
 N^{-1/3} K_1^{2/3} + \cdots$ 
& $\displaystyle E = k R_1^{1/2} N^{-1/2} K_1^{1/2} + \cdots$  \\ \hline 
\rule[-5mm]{0mm}{13mm} I$_\B$, $\w = 0$      & $\displaystyle E -
  \frac{K_2}{R_0} = - k \frac{N^2 K^3_2}{R_0^7} +
  \cdots $ & $\displaystyle E = k R_1^{1/2} N^{-1/2} K_2^{1/2} + \cdots$
 \\ \hline
\rule[-5mm]{0mm}{13mm} II$_\B$, $\nu_2 = 0$  & $\displaystyle E = k N^{2/3}
  K_3^{2/3} + \cdots$ & 
$\displaystyle E - \frac{N K_3}{R_1} = k R_1 N^{1/3} K_3^{1/3} + \cdots$   \\ \hline
\rule[-5mm]{0mm}{13mm} II$_\B$, $\nu_3 = 0$  & $\displaystyle E -
  \frac{K_2}{R_0} = - k \frac{K^3_2}{R_0^7} +
  \cdots $  &
$\displaystyle E = k K_2^{2/3} + \cdots$   \\ \hline
\rule[-5mm]{0mm}{13mm} III$_\B$, $\nu_3 = 0$ & $\displaystyle E = k R_0^{1/2} K_1^{1/2} + \cdots$ &
$\displaystyle E = k K_1^{2/3} + \cdots$  \\ \hline
\rule[-5mm]{0mm}{13mm} III$_\B$, $\w = 0$    & $\displaystyle E = k
  R_0^{1/2} N^{1/2} K_3^{1/2} + \cdots$  &
$\displaystyle E - \frac{N K_3}{R_1} = k R_1 N^{1/3} K_3^{1/3} + \cdots$   \\ \hline
  \end{tabular}
\end{table}

The results in Table 2 have a physical interpretation. Note that there
are four types of leading order behaviour. Use $K$ to denote a generic charge and
$R$ to denote either $R_1$ or $R_2$.

\begin{itemize}

\item $\displaystyle E = k R^{1/2} K^{1/2}$: This is the well known
Regge relation for strings in flat space. It arises when the
$\d$-direction of the membrane is wrapped on a stabilised $U(1)$ and
when the direction of rotation is a $U(1)$ that is not stabilised
(i.e. collapsing if we are at the origin or expanding if we are going
to infinity).

\item $\displaystyle E = k K^{2/3}$: This is the result for membranes
rotating in flat space. It arises when neither the $\d$-direction nor
the direction of rotation is stabilised.

\item $\displaystyle E - \frac{K}{R} = k R K^{1/3}$: This result
arises for long strings when the $\d$-direction is not stabilised, but the
direction of rotation is stabilised. Interestingly, this relation was
also observed in a different configuration \cite{Alishahiha:2002sy} in
$AdS_7\times S^4$, suggesting perhaps that it is quite generic.

\item $\displaystyle E - \frac{K}{R} = - k \frac{K^3}{R^7}$: This
result arises for short strings when the $\d$-direction collapses, but the
direction of rotation does not collapse.

\end{itemize}

The behaviour of the energy-charge relationship would thus seem to
depend on whether the wrapped circle and the circle of rotation are
stabilised. In the above configurations one case is missing, there is
no case in which both the $\d$-circle and the circle of rotation do
not collapse. For short strings, we will find such a configuration in
the ${\Bbb D}_7$ metrics below, because more circles are non-shrinking
at the origin. However, within the set of configurations we have
considered thus far, we cannot find a configuration in which two
circles are stabilised at infinity, because the $G_2$ metrics only
have one stabilised circle. We might expect such a configuration to
give logarithms by analogy with the previous section when we considered
membranes rotating on $AdS_4\times S^7$: to move from a relationship
of the form $E-K=K^{1/3}$ to a relationship $E-K=\ln K$, we changed
the wrapped circle to make it stabilised. To achieve this in the
present case, we need to use the non-compact directions. However, due
to the tension of the membrane, one cannot simply have a closed
membrane in flat space. The resolution is to consider an infinite
membrane in the non-compact directions and study the energy density of
the configuration. Insofar as the equations are concerned, this is
effectively the same as wrapping the membane.

\subsubsection{Using the non-compact directions: logarithms}

Writing the eleven dimensional metric as
\be
\frac{1}{l_{11}^2} ds^2_{11} = -dt^2 + dx^2 + dy^2 + dz^2 + ds^2_7 , 
\ee
the following configuration, which we denote IV$_{\B}$, has the
desired feature of having both a wrapped and a rotating direction
asymptotically stabilised. The nontrivial coordiantes are
\be
t = \k \t, \;\; \psi_3 = \nu_3 \t, \;\;  r = r(\s), \;\;
x = \lambda \d .
\ee
The remaining coordinates are trivial
\be
\phi_3 = \phi_4 = 0, \;\; y = z = 0, \;\; \q_1 = \q_2 =
\p/2, \;\; \psi_4 = 0 .
\ee
One might also want to consider having the rotation in the non-compact
direction, but this seems to cause difficulties with the
implementation of the endpoint constraint (\ref{eq:endpoint}).
The target space metric seen by the membrane now becomes
\be
\frac{1}{l_{11}^2} ds^2_{M2} = -dt^2 + dr^2 + \frac{c(r)^2}{N^2}
d\psi_3^2 + d x^2 .
\ee
The action, energy and charge, {\it per unit length} along the
non-compact $x$ direction, are easily worked out to be
\begin{itemize}
\item Case IV$_\B$

\bea
I & = & \frac{-8\p}{(2\p)^2}\int d\t\int_0^{r_0} dr
\sqrt{\k^2 - \nu_3^2 c(r)^2/N^2} , \\
E & = & \frac{8\p}{(2\p)^2}\int_0^{r_0} dr
\frac{|c(r_0)|}{\sqrt{
\left[c(r_0)^2-c(r)^2\right]}} , \\
K_3 & = & \frac{8 \p}{N (2\p)^2}\int_0^{r_0} dr
\frac{c(r)^2}{\sqrt{
\left[c(r_0)^2-c(r)^2\right]}} .
\eea

\end{itemize}
In the large membrane limit, $r_0\to\infty$, these integrals are
dominated by their behaviour at $r_0$, thus we may expand the integrand
and evaluate only at $r_0$. We substitute the expansion of $c(r)$ to
second order into the integrand and evaluate the integral. Expanding
the integrand fully before integrating will not give the correct
answer, as no logs will appear.

The integrals give, to leading and subleading order
\bea
K_3 = k \frac{-i r_0 R_1}{2 N} \left[ 3 K(\sqrt{2r_0^2/(9 R_1^2)-1})
+ E(\sqrt{2r_0^2/(9 R_1^2)-1}) \right]  , \nonumber \\
E = k \frac{i r_0}{2} \left[ K(\sqrt{2r_0^2/(9 R_1^2)-1}) - E(\sqrt{2r_0^2/(9 R_1^2)-1}) \right] ,
\eea
Where $K(x)$ and $E(x)$ are complete Elliptic integrals of the first
and second kind. The constants $k$ in the above two lines are equal,
but below we use $k$ to denote any constant, with dependence on $\r_0$
and $R_1$ kept explicit.

In order to evaluate these integrals we need the following expressions
for the asymptotics as $x \to \infty$ of the Elliptic integrals
\bea
K(\sqrt{x-1}) \sim - \frac{i}{2} x^{-1/2} (\ln x + i \pi) , \nonumber
\\
E(\sqrt{x-1}) \sim i x^{1/2} .
\eea
These formulae are easily derived by expressing the complete elliptic integrals
as hypergeometric functions and then using the Pfaff and Gauss theorems for
hypergeometric functions \cite{book}.

Thus we have that whilst
\be
K_3 = \frac{k r_0^2}{N} + \cdots ,
\ee
the difference
\be
E - \frac{N K_3}{R_1} = k R_1 \ln \frac{r_0}{R_1} + \cdots .
\ee
Combining these two expressions gives the new kind of behaviour for
long membranes
\be
E - \frac{N K_3}{R_1} = k R_1 \ln \frac{N K_3}{R_1^2} + \cdots .
\ee
This behaviour is different from the behaviours of the previous
section because both the direction of wrapping and the direction of
rotation are stabilised as we go to infinity.

For short membranes with this configuration, we get $\displaystyle E =
k N^{1/2} K_3^{1/2} + \cdots$ as expected for a membrane where the
$\d$-direction is stabilised but the rotation direction collapses at
the origin. These solutions thus realise a transition from Regge behaviour for short membranes,
to logarithmic behaviour for long membranes without finite size effects \cite{Armoni:2002fr}.

Another way of getting around the fact that a closed tubular membrane
in flat space can't exist as a static solution due to the membrane
tension is to consider a pulsating membrane solution, analogous to the
well-known pulsating closed string solution. The solution will not be
particularly straightforward to construct in the present context. Such
a solution may have a tunable amplitude, in which case one could make the
energy of the pulsations negligible compared with the energy of the
rotation and therefore the calculations of this section will go
through as the dominant effect.

\subsection{Membranes on the ${\Bbb D}_7$ family}

\subsubsection{Metric formulae}

The metrics can be written in the form
\bea
ds^2_7 = dr^2 + a(r)^2\left[(\S_1 + g(r)\s_1)^2
+(\S_2 + g(r)\s_2)^2\right]+ c(r)^2(\S_3 + g_3(r) \s_3)^2  \nn \\
+ b(r)^2\left[\s_1^2 + \s_2^2 \right] + f(r)^2 \s_3^2 ,
\eea
where $\S_i,\s_i$ are left-invariant one-forms on the $SU(2)$s, as previously.
The six functions are not all independent
\be
g(r) = \frac{-a(r)f(r)}{2b(r)c(r)} , \;\; g_3(r) = -1 + 2g(r)^2 .
\ee
None of the radial functions are known explicitly, although the
asymptotics at the origin and at infinity are known. The asymptotics
are found by finding Taylor series solutions to the first order equations
for the radial functions. The equations are \cite{Cvetic:2001ih}
\bea\label{eq:Deqns}
\dot{a} = -\frac{c}{2a} + \frac{a^5 f^2}{8 b^4 c^3}, & \;\; &
\dot{b} = -\frac{c}{2b} - \frac{a^2 (a^2-3c^2)f^2}{8b^3c^3}, \nn \\
\dot{c} = -1+\frac{c^2}{2 a^2}+\frac{c^2}{2 b^2}-\frac{3 a^2
f^2}{8b^4}, & \;\; &
\dot{f} = -\frac{a^4 f^3}{4 b^4 c^3}.
\eea
As $r\to 0$ one has
\bea\label{eq:Dat0}
a(r) & = & \frac{r}{2}-\frac{(q_0^2+2)r^3}{288 R_0^2} -
\frac{(-74-29q_0^2+31q_0^4)r^5}{69120 R_0^4} + \cdots  , \nn \\
b(r) & = & R_0 - \frac{(q_0^2-2)r^2}{16 R_0} -
\frac{(13-21q_0^2+11q_0^4) r^4}{1152 R_0^3} + \cdots  , \nn \\
c(r) & = & -\frac{r}{2} - \frac{(5q_0^2-8)r^3}{288 R_0^2} -
\frac{(232-353q_0^2+157q_0^4) r^5}{34560 R_0^4}+ \cdots , \nn \\
f(r) & = & q_0 R_0 + \frac{q_0^3 r^2}{16 R_0} +
\frac{q_0^3(-14+11q_0^2) r^4}{1152 R_0^3} + \cdots ,
\eea
where $q_0$ and $R_0$ are constants.
Note that $a(r)$ and $c(r)$ collapse and the other two functions do not.
As $r\to\infty$ we have
\bea\label{eq:Datinfinity}
a(r) & = &  \frac{r}{\sqrt{6}} - \frac{\sqrt{3} q_1 R_1}{\sqrt{2}} +
\frac{(27\sqrt{6}- 96 h_1 ) R_1^2}{96r} + \cdots , \nn \\
b(r) & = &  \frac{r}{\sqrt{6}} - \frac{\sqrt{3} q_1 R_1}{\sqrt{2}} +
\frac{h_1 R_1^2}{r} + \cdots , \nn \\
c(r) & = &  \frac{-r}{3} + q_1 R_1 - \frac{9 R_1^2}{8r} + \cdots , \nn \\
f(r) & = &  R_1 - \frac{27 R_1^3}{8 r^2} - \frac{81 R_1^4 q_1}{4 r^3} + \cdots .
\eea
With constants $R_1,q_1,h_1$. Note that $f(r)$ stabilises.
Three constants appear to this order, whilst there were only two
constants in the expansion about the origin. This just means that for
some values of these constants, the corresponding solution will
diverge before it reaches zero. In any case, we find no $h_1$
dependence in the results below.

\subsubsection{Membrane configurations}

The situation is essentially the same as for the $\B_7$ family of
metrics. Again one has three commuting $U(1)$ isometries, generated by
$\del_{\phi_1} \subset SU(2)_L^{\s}$, $\del_{\phi_2}
\subset SU(2)_L^{\S}$ and $\del_{\psi_1}+\del_{\psi_2} \subset
SU(2)_R^D$. One difference, however, is that now $\del_{\phi_1}$
generates a $U(1)$ that does not collapse and $\del_{\phi_2}$
generates a circle that does collapse, so there is no need to change
variables to $\phi_3$ and $\phi_4$ as previously. In fact, such a
change would not give a valid solution. We do however need to define
\be
\psi_3 = \psi_1 + \psi_2, \;\; \psi_4 = \psi_1 - \psi_2 .
\ee
Note that now $\psi_3,\psi_4$ have ranges of $8\p$ whilst
$\phi_1,\phi_2$ have ranges of $2\pi$.
Three commuting $U(1)$ isometries are then $\del_{\phi_1}$,
$\del_{\phi_2}$ and $\del_{\psi_3}$. There are no branes in reduction
of these configurations to well-defined IIA solutions. The circle generated by
$\del_{\phi_1}$ and $\del_{\psi_3}$ do not collapse in the interior
and thus rotation in these directions corresponds to charges, $K_1$
and $K_2$ respectively. The $\del_{\phi_2}$ circle does collapse and
rotation about this circle will give a charge denoted by $K_3$.

Similar to before, we take
\be
x = y = z = 0, \;\; \q_1 = \q_2 = \p/2, \;\; \psi_4 = \p/2 .
\ee
Note that the value of $\psi_4$ is different. This value is needed to
diagonalise the metric seen by the membrane and hence satisfy the constraints.
As in the previous subsection, there are three possible configurations for the nontrivial
directions, shown in Table 3.

\begin{table}[h]
{\bf Table 3: } Three rotating membrane configurations on the ${\Bbb D}_7$ metrics \\
  \begin{tabular}{|c|c|c|c|} \hline
Target space coordinate & Configuration I$_{\Bbb D}$ & Configuration II$_{\Bbb D}$ &
  Configuration III$_{\Bbb D}$ \\ \hline \hline
$\phi_1 =$ & $\w_1 \t$ & $\l \d$    & $\w_1 \t$ \\ \hline
$\phi_2 =$ & $\nu \t$    & $\nu \t$    & $\l \d$ \\ \hline
$\psi_3 =$ & $\l \d$    & $\w_2 \t$ & $\w_2 \t$ \\ \hline
$t =$ & \multicolumn{3}{|c|}{$\k \t$} \\ \hline
$r =$ &  \multicolumn{3}{|c|}{$r(\s)$} \\ \hline
  \end{tabular}
\end{table}

The target space metric that is seen by the membrane, after doing the
${\mathbb{Z}}_N$ quotient on $\psi_3$, is
\bea
\frac{1}{l_{11}^2} ds^2_{M2} & = & -dt^2 + dr^2 + \left[\frac{1}{4} f(r)^2 + c(r)^2 g(r)^4
\right] \frac{d\psi_3^2}{N^2} + \left[a(r)^2 g(r)^2 + b(r)^2\right] d\phi_1^2 +
a(r)^2 d\phi_2^2 \nn \\
& \equiv & -dt^2 + dr^2 + \frac{C(r)^2}{N^2} d\psi_3^2 + B(r)^2 d\phi_1^2 +
A(r)^2 d\phi_2^2 ,
\eea
where we have introduced functions $A(r), B(r), C(r)$ in order to make
the following formulae less ugly. The asymptotics for these functions
follow from the limits (\ref{eq:Dat0}) and (\ref{eq:Datinfinity}) and
the algebraic equations in (\ref{eq:Deqns}). As $r\to 0$ we have
\bea
A(r) & = & \frac{r}{2} - \frac{(q_0^2+2)r^3}{288 R_0^2} -
\frac{(-74-29q^2+31q^4)r^5}{69120 R^4} +  \cdots  , \nn \\
B(r) & = & R_0 + \frac{(4-q_0^2)r^2}{32 R_0} -
\frac{(61q^4-152q^2+208)r^4}{18432 R^3} + \cdots  , \nn \\
C(r) & = & \frac{q_0 R_0}{2} + \frac{3q_0^3 r^2}{64 R_0} -
\frac{(-160+121q^2)q^3r^4}{12288 R^3} + \cdots , 
\eea
and as $r\to\infty$ we have
\bea
A(r) & = & \frac{r}{\sqrt{6}} - \frac{\sqrt{3} q_1 R_1}{\sqrt{2}} + \frac{(27\sqrt{6} - 96 h_1)R_1^2}{96 r} + \cdots  , \nn \\
B(r) & = & \frac{r}{\sqrt{6}} - \frac{\sqrt{3} q_1 R_1}{\sqrt{2}} +
\frac{(18\sqrt{6} + 96 h_1)R_1^2}{96 r} + \cdots  , \nn \\
C(r) & = & \frac{R_1}{2} - \frac{9 R_1^3}{8 r^2} - \frac{27 q_1
R_1^4}{4 r^3}  + \cdots .
\eea
The only nontrivial constraint from (\ref{eq:constraints})
now implies that
\be
\left(\frac{d r}{d \s}\right)^2 =
\left\{
\begin{array}{r}
\frac{\k^2 - B(r)^2 \w_1^2 - A(r)^2 \nu^2}{C(r)^2 l_{11}^2 / N^2} \quad\quad (\mbox{Case I$_{\Bbb D}$}) \\
\frac{\k^2 - A(r)^2 \nu^2 - C(r)^2 \w_2^2 / N^2}{B(r)^2 l_{11}^2}  \quad\quad (\mbox{Case II$_{\Bbb D}$}) \\
\frac{\k^2 - B(r)^2 \w_1^2 - C(r)^2 \w_2^2 / N^2}{A(r)^2 l_{11}^2} \quad\quad (\mbox{Case III$_{\Bbb D}$})
\end{array}
\right.
\ee
The issue of consistency is more subtle in these cases than in the
$\B_7$ cases. This is because there are more cross terms in the
metric. It turns out that in order for the ans\"atze to be consistent,
one can only have one of the angular momenta to be nonzero in any
given solution. Thus, for example, in the type I$_{\Bbb D}$ solution one
must have either $\w_1 = 0$ or $\nu = 0$ in order for the
configuration to solve the equations of motion. These are the
configurations we shall consider below.

\subsubsection{Energy and other conserved charges}

Again, the differing numerical factors in the expressions below are 
due to differences in the ranges of angles.

\begin{itemize}

\item Case I$_{\Bbb D}$

\bea
I & = & \frac{-32\p}{N (2\p)^2}\int d\t\int_0^{r_0} dr |C(r)| \sqrt{\k^2 - B(r)^2
\w_1^2 - A(r)^2 \nu^2} , \\
E & = & \frac{32\p}{N (2\p)^2}\int_0^{r_0} dr
\frac{|C(r)|\sqrt{\w_1^2 B(r_0)^2 + \nu^2
A(r_0)^2}}{\sqrt{\w_1^2\left[B(r_0)^2-B(r)^2\right] + \nu^2
\left[A(r_0)^2-A(r)^2\right]}} , \\
K_1 & = & \frac{32\p}{N (2\p)^2}\int_0^{r_0} dr
\frac{\w_1 |C(r)| B(r)^2}{\sqrt{\w_1^2\left[B(r_0)^2-B(r)^2\right] + \nu^2
\left[A(r_0)^2-A(r)^2\right]}}, \\
K_2 & = & 0, \\
K_3 & = & \frac{32\p}{N (2\p)^2}\int_0^{r_0} dr
\frac{\nu |C(r)| A(r)^2}{\sqrt{\w_1^2\left[B(r_0)^2-B(r)^2\right] + \nu^2
\left[A(r_0)^2-A(r)^2\right]}}.
\eea

\item Case II$_{\Bbb D}$

\bea
I & = & \frac{-8\p}{(2\p)^2}\int d\t\int_0^{r_0} dr
|B(r)|\sqrt{\k^2 - A(r)^2 \nu^2 - \w_2^2 C(r)^2 / N^2} , \\
E & = & \frac{8\p}{(2\p)^2}\int_0^{r_0} dr
\frac{|B(r)|\sqrt{\nu^2 A(r_0)^2 + \w_2^2
C(r_0)^2 / N^2}}{\sqrt{\nu^2\left[A(r_0)^2-A(r)^2\right] + \w_2^2
\left[C(r_0)^2-C(r)^2\right] / N^2}}, \\
K_1 & = & 0, \\
K_2 & = & \frac{8\p}{N^2 (2\p)^2}\int_0^{r_0} dr
\frac{\w_2 |B(r)| C(r)^2}{\sqrt{\nu^2\left[A(r_0)^2-A(r)^2\right] + \w_2^2
\left[C(r_0)^2-C(r)^2\right] / N^2}}, \\
K_3 & = &  \frac{8\p}{(2\p)^2}\int_0^{r_0} dr
\frac{\nu |B(r)| A(r)^2}{\sqrt{\nu^2\left[A(r_0)^2-A(r)^2\right] + \w_2^2
\left[C(r_0)^2-C(r)^2\right] / N^2}}.
\eea

\item Case III$_{\Bbb D}$

\bea
I & = & \frac{-8\p}{(2\p)^2}\int d\t\int_0^{r_0} dr |A(r)|
\sqrt{\k^2 - B(r)^2 \w_1^2 - \w_2^2 C(r)^2/N^2} , \\
E & = & \frac{8\p}{(2\p)^2}\int_0^{r_0} dr
\frac{|A(r)|\sqrt{\w_1^2 B(r_0)^2 + \w_2^2
C(r_0)^2/N^2}}{\sqrt{\w_1^2\left[B(r_0)^2-B(r)^2\right] + \w_2^2
\left[C(r_0)^2-C(r)^2\right]/N^2}}, \\
K_1 & = & \frac{8\p}{(2\p)^2}\int_0^{r_0} dr
\frac{\w_1 |A(r)| B(r)^2}{\sqrt{\w_1^2\left[B(r_0)^2-B(r)^2\right] + \w_2^2
\left[C(r_0)^2-C(r)^2\right]/N^2}} , \\
K_2 & = & \frac{8\p}{N^2 (2\p)^2}\int_0^{r_0} dr
\frac{\w_2 |A(r)| C(r)^2}{\sqrt{\w_1^2\left[B(r_0)^2-B(r)^2\right] + \w_2^2
\left[C(r_0)^2-C(r)^2\right]/N^2}} , \\
K_3 & = & 0 .
\eea

\end{itemize}
These integrals are now performed in the small and large membrane
limits in the same way as for the $\B_7$ metrics. The results are
presented in Table 4. Again, $k$ denotes positive numerical constants, with
dependence on $R_0,R_1,q_0,q_1,N$ kept explicit. It is not surprising
that a dependence on $q_0$ now emerges because, unlike the $\B_7$
cases, the principle interpretation of this parameter is as measuring
the squashing of the bolt at the origin \cite{Cvetic:2001ih}.

\begin{table}[h]
{\bf Table 4: } Energy - Charge relations for membranes on ${\Bbb D}_7$ metrics \\
  \begin{tabular}{|c|c|c|c|} \hline
Configuration & $r_0 \to 0$ & $r_0 \to \infty$ \\ \hline \hline
\rule[-5mm]{0mm}{13mm} I$_{\Bbb D}$, $\w_1 = 0$   & $\displaystyle E =
  k R_0^{1/2} q_0^{1/2} N^{-1/2} K_3^{1/2} + \cdots$ 
& $\displaystyle E = k R_1^{1/2} N^{-1/2} K_3^{1/2} + \cdots$   \\ \hline
\rule[-5mm]{0mm}{13mm} I$_{\Bbb D}$, $\nu = 0$    & $\displaystyle E = \frac{K_1}{R_0} = k R_0^2
  \frac{q_0}{N\sqrt{4-q_0^2}} + \cdots$ 
& $\displaystyle E = k R_1^{1/2} N^{-1/2} K_1^{1/2} + \cdots$    \\ \hline
\rule[-5mm]{0mm}{13mm} II$_{\Bbb D}$, $\w_2 = 0$  & $\displaystyle E = k R_0^{1/2} K_3^{1/2} + \cdots $ 
& $\displaystyle E = k K_3^{2/3} + \cdots$   \\ \hline
\rule[-5mm]{0mm}{13mm} II$_{\Bbb D}$, $\nu = 0$   & $\displaystyle E =
  \frac{2 N K_2}{R_0 q_0} = k \frac{R_0^2}{q_0} + \cdots$
& $\displaystyle E - \frac{2 N K_2}{R_1} = k R_1 N^{1/3} K_2^{1/3} + \cdots$  \\ \hline
\rule[-5mm]{0mm}{13mm} III$_{\Bbb D}$, $\w_2 = 0$ & $\displaystyle E -
  \frac{K_1}{R_0} = - k\frac{(4-q_0^2)^2 K_1^3}{R_0^7} + \cdots$ 
& $\displaystyle E = k K_1^{2/3} + \cdots$   \\ \hline
\rule[-5mm]{0mm}{13mm} III$_{\Bbb D}$, $\w_1 = 0$ & $\displaystyle E -
  \frac{2 N K_2}{q_0 R_0} = - k \frac{q_0
  N^3 K_2^3}{R_0^7} + \cdots$ 
& $\displaystyle E - \frac{2 N K_2}{R_1} = k R_1 N^{1/3} K_2^{1/3} + \cdots$   \\ \hline
  \end{tabular}
\end{table}

The behaviours observed are the same as for the $\B_7$ metrics, except
that there is one new possibility for short strings

\begin{itemize}

\item $\displaystyle E = \frac{K}{R} = k R^2$: This arises
when the $\d$-direction does not collapse at the origin, so the
rotation is string-like, and the direction of rotation also does not
collapse. There will be a dependence on $q$.

\end{itemize}

\subsubsection{Using the non-compact directions again: logarithms}

Writing the eleven dimensional metric as
\be
\frac{1}{l_{11}^2} ds^2_{11} = -dt^2 + dx^2 + dy^2 + dz^2 + ds^2_7 , 
\ee
we may do exactly the same calcutions as we did before for the $\B_7$ metrics.
The configuration that follows will be denoted IV$_{\Bbb D}$,
\bea
t = \k \t, \;\; \psi_3 = \w_2 \t, \;\; \phi_1 = \phi_2 = 0, \;\; r = r(\s),
\nonumber \\
y = z = 0, \;\;  x = 0, \;\; \phi = \lambda \d, \;\; \q_1 = \q_2 = \p/2, \;\; \psi_4 = \p/2 .
\eea
The target space metric seen by the membrane now becomes
\be
\frac{1}{l_{11}^2} ds^2_{M2} = -dt^2 + dr^2 + \frac{C(r)^2}{N^2} d\psi_3^2 + dx^2 .
\ee
The action, energy and charge, per unit length along the $x$ direction, are easily worked out to be
\begin{itemize}
\item Case IV$_{\Bbb D}$

\bea
I & = & \frac{-8\p}{(2\p)^2}\int d\t\int_0^{r_0} dr
\sqrt{\k^2 - \w_2^2 C(r)^2/N^2} , \\
E & = & \frac{8 \p}{(2\p)^2}\int_0^{r_0} dr
\frac{|C(r_0)|}{\sqrt{
\left[C(r_0)^2-C(r)^2\right]}} , \\
K_3 & = & \frac{8\p}{N (2\p)^2}\int_0^{r_0} dr
\frac{C(r)^2}{\sqrt{
\left[C(r_0)^2-C(r)^2\right]}} .
\eea

\end{itemize}
Obviously, this is exactly the same as the $\B_7$ case but with $c(r)
\to C(r)$. The asymptotic expansions to second order are the same for
$c(r)$ and $C(r)$ if we let $R_1 \to R_1/2$. Thus we get the same result
\be
E - \frac{2 N K_3}{R_1} = k R_1 \ln \frac{N K_3}{R_1^2} + \cdots .
\ee

For short membranes with this configuration, we get $\displaystyle E =
2 N K_3 /(R_1 q_1) = k R_1/q + \cdots$. As commented before, we might
expect to also find a pulsating membrane solution with these energy -
charge relations.

\section{Discussion, comments regarding dual operators and open issues}

Let us first review what we have done and found in this work.
Motivated by the recent developments mentioned in the first section, we 
studied membranes rotating in different geometries that are of interest as
duals to gauge theories.

To start with, we considered an $AdS_4 \times M_7$ spacetime.
Depending on the holonomy of $M_7$, these are dual to $2+1$ dimensional
conformal field theories with a varying number of preserved 
supersymmetries. In all of these manifolds, we found that rotating 
membrane configurations may develop relations for the energy E, spin S, and 
R-symmetry angular momentum J, of the form $E-S \sim \ln S$ and 
$E-J-S \sim 1/J \ln^2(S/J)$, as had previously been found for strings
on various backgrounds. The same logarithmic results were found for 
membranes moving in a warped $AdS_5\times M_6$ geometry that is dual to
a four dimensional $N=2$ conformal field theory. We also recovered
previous non-logarithmic results for membranes and explained the
difference between the logarithmic and non-logarithmic cases in terms
of whether the direction
wrapped by the membrane was stabilised at infinity or not.

According to the correspondence between high angular momentum
strings/membranes and `long' operators \cite{Gubser:2002tv},
these rotating membranes should be dual to certain twist two operators in
the corresponding conformal field theory that have anomalous dimensions
given by the relation between energy (or conformal dimension), spin
and J-charge calculated on the gravity side of the duality.
These results point to the fact that for geometries of the form $AdS_p 
\times M_q$, it will be possible to find membrane/string configurations dual
to `long' twist two operators.

Given these results, we were lead to a very natural extension; geometries 
that are not of the form $AdS_p\times M_q$.
Section four of this work presents a very detailed study of membranes 
rotating in M-theory backgrounds of the form ${\mathbb{R}}^{1,3}
\times M_7$, where $M_7$ is now a non-compact $G_2$ holonomy manifold. These
backgrounds are thought to be dual to ${\mathcal{N}}=1$ SYM, which is a
confining `QCD-like' theory.

The results of section four could be summarised as follows. We have
found rotating membrane
configurations that should be dual to operators with energy-angular 
momentum relations, using $K$ to denote the angular momentum/dual charge,
of the following form for small quantum numbers
$E \sim K^{2/3},\;\; E- K\sim K^3,\;\; E\sim K^{1/2},\;\; 
E\sim K=\mbox{constant}$. When continued
to the large quantum number regime these may become $E\sim K^{1/2},\;\; E\sim 
K^{2/3},\;\; E- K\sim K^{1/3},\;\; E- K\sim \ln K.$
For the logarithmic cases, we considered energy and charge densities of a
noncompact membrane.
Some of these configurations seem to realise the proposal of 
\cite{Armoni:2002fr} for rotating solutions in a confining geometry
to exhibit a transition for Regge-like to D.I.S.-like behaviour without
finite size effects.

Several comments are in order. First of all, we consider these results to be 
interesting. Not many dynamical or quantitative tests of the duality 
between M theory on $G_2$ manifolds and ${\mathcal{N}}=1$ SYM theory
seem to exist. We hope that our results are a step towards an
understanding of the duality that involves both a dynamical and 
a quantitative statement. Indeed, the fact that we obtained results
that look very much like they should correspond to anomalous
dimensions of operators, suggests that the energy of gravity states corresponds to
the dimension of gauge theory operators. This is not at all otherwise obvious,
given the lack of conformal symmetry and the lack of a holographic formulation of
the duality that explicitly links bulk states with boundary operators.

We should point out that our results leave many issues open. These
issues seem to be inextricably tied up with limitations of current
understanding of the 
duality. To begin with, due to the running of the gauge theory dual
coupling, it is not evident how to read off
from our solutions the 't Hooft coupling dependence of the relation 
between energy/dimension and spin/charge in the field theory. Then,
the interpretation of the field theory dual charge
to the angular momenta of the membrane is not totally clear. Given
that the rotations are not in the four flat non-compact directions of
the spacetime, it is
not obvious why the charge should be four dimensional Lorentzian spin and if
it is not spin, then it is not clear what else it
could be. As well as the known behaviours, like $E\sim 
K^{1/2},\;\; E\sim K^{2/3}$, which are Regge-like behaviours for short membranes,
and $E-K\sim \ln K$, which is the D.I.S/twist two-like behaviour,
we obtained other 
relations such as those of the form $E-K\sim K^{1/3},\;\; E-K\sim K^3$.
The first type of relation appeared previously for strings moving in Witten's QCD confining model 
\cite{Armoni:2002fr}. The second type does not seem to have been
previously studied. It is not clear which `QCD-like' 
operator will be dual to these last two configurations. We must keep in 
mind that, in the gravity approximation, M-theory on a $G_2$ manifold
is not dual to pure ${\mathcal{N}}=1$ SYM. Indeed Kaluza-Klein and bulk modes are not decoupled 
from the $3+1$ gauge theory, a feature that seems to afflict any study 
involving D6 branes.

One might speculate that the logarithmic configurations we found on the $G_2$
spacetimes could be related to large Lorentzian spin operators in field theory
via Wilson lines. Wilson lines are closely related to the twist two
operators of form (\ref{eq:theop}), see for example
\cite{Kruczenski:2002fb} and references therein. The membrane
configurations in question, called IV$_{\B}$ and IV$_{\Bbb D}$ above,
form an infinite line in the noncompact directions. Some related comments were made in
\cite{Sezgin:2002rt}.

Finally, in the following appendices, we set up a formalism to study certain
string configurations on warped AdS backgrounds, that are general
confining backgrounds. We then end by explicitly checking
non-supersymmetry of the membrane configurations we considered on
$G_2$ backgrounds.

\vspace{1.5cm}
\centerline{\bf Acknowledgements}

We would like to thank Robert Thorne for a very useful discussion and
Gregory Korchemsky for very extensive and useful e-mail correspondence.
We would also like to thank Gary Gibbons,
Rub\'en Portugues and in particular Nemani Suryanarayana for
discussions. SAH is funded by the Sims
scholarship and CN is funded by PPARC.

\appendix

\section{Strings moving in warped AdS spaces}

In this section we will consider strings rotating in the background 
geometry generated by a D4-D8 system. The objective is to study
effects of warp factors on the rotating configurations. The subsection below takes a
more general approach. The geometry is given in 
\cite{Brandhuber:1999np}
 and reads in Einstein frame,
\beq
ds^2= (\sin\xi)^{1/12} [- dt^2 \cosh^2\rho + d\rho^2 + \sinh^2\rho 
d\Omega_4^2 + \frac{2}{g^2} (d\xi^2 + \frac{\cos^2\xi}{4} (d\theta^2 + 
d\phi^2 + d\psi^2+ 2 \cos\theta d\phi d\psi) )] ,
\label{metricwarped}
\eeq
where $g$ is a constant and
\beq
d\Omega_4^2 = d\theta_1 + \cos^2\theta_1 d\theta_2 + 
\cos^2\theta_1\cos^2\theta_2 d\theta_3^2 + \cos^2\theta_1\cos^2\theta_2 
\cos^2\theta_3 d\theta_4^2 ,
\eeq
and the matter fields are
\beq
e^{-6/5\phi}=\sin\xi,\;\;\; F_{4}= \sin^{1/3}\xi \cos^3\xi
\mbox{vol}(S^4) .
\eeq
We will consider here a configuration given by
\beq
\rho=\rho(\sigma),\;\; t= \kappa \tau \;\; \theta_4 = \omega \tau,\;\; \xi= 
\xi(\sigma), \;\; 
\psi= \sqrt{2}\nu \tau, \;\; \theta_i=0,\;\; \theta=0 .
\eeq
Plug this configuration into the string action, and neglect 
the term that is proportional to the dilaton, since it is an $\alpha'$ correction 
and we are not considering in the metric (\ref{metricwarped}) any stringy 
corrections. We have an action
\beq
S=\frac{2P}{g^2}\int d\sigma f(\xi) \left[ \xi'^2 + \frac{g^2}{2}\rho'^2 
+ \frac{1}{2} \left(g^2 \kappa^2 \cosh^2\rho - \omega^2 g^2 
\sinh^2\rho\right) - \nu^2 \cos^2\xi \right],
\label{actionwarped}
\eeq
with $f(\xi)= \sin^{1/12}\xi(\sigma)$ and ${}^{\prime}$ denoting $d/d{\sigma}$.
The equations of motion are 
\beq
\frac{d}{d\sigma}(f(\xi) \rho') - f(\xi) (\omega^2 - \kappa^2)\cosh\rho
\sinh\rho=0 ,
\label{eqwarped1}
\eeq
\beq
\frac{d}{d\sigma}(f(\xi) \xi')= \frac{g^2}{4}\frac{df(\xi)}{d\xi} (\kappa^2 \cosh^2\rho - \omega^2 
\sinh^2\rho - \frac{2 \nu^2 \cos^2 \xi}{g^2} + \frac{2\xi'^2}{g^2} + \rho'^2) + f(\xi)\nu^2\cos\xi 
\sin\xi ,
\label{eqwarped2}
\eeq
and the constraint reads
\beq
f(\xi) \left(\frac{2\xi'^2}{g^2} + \rho'^2 - \kappa^2 \cosh^2\rho + \omega^2 
\sinh^2\rho + \frac{2 \nu^2 \cos^2 \xi}{g^2}\right) = 0 .
\label{constraintwarped}
\eeq
One can check that the derivative of 
the constraint can be split up to give the second order equations of motion.

We can consider special cases of the previous configurations. 
In the case
where the warping angle $\xi$ is taken to be a 
constant $\xi_* = \pi/2$. Note that this specific value is needed to
solve the equations of motion. There is no R-charge in this
configuration. The computations are very similar to the 
cases 
analysed in \cite{Gubser:2002tv} and we get the same result 
with operators satisfying $E-S \sim \ln S$. 
Besides, one can consider the case in which the coordinate $\rho=\rho_*=0$ is 
constant, again with a value specified by the equations of motion. In
this case, the equations of motion reduce to
\beq
\xi'' = \nu^2 \cos\xi \sin\xi, \;\;\; \xi'^2 + \nu^2 \cos^2\xi= 
\frac{g^4}{4} \kappa^2 ,
\eeq
The turning point will be
\beq
\cos^2\xi_0= \frac{g^4}{4\nu^2} \kappa^2 .
\eeq
We can compute the energy and angular momentum for this 
configuration to be given by
\beq
E= \frac{2}{g^2}P \kappa \int^\xi_0 
\frac{\sin^{1/12}\xi}{\sqrt{\frac{g^4}{4\nu^2} \kappa^2 
 - \nu^2 \cos^2\xi}} d\xi ,
\label{ewarped}
\eeq
\beq
J= \frac{2}{g^2}P \nu \int^\xi_0
\frac{\sin^{1/12}\xi}{\sqrt{\frac{g^4}{4\nu^2} \kappa^2 
 - \nu^2 \cos^2\xi}} \cos^2\xi d\xi .
\label{jwarped}
\eeq
As we have done in previous sections, we expand the integrals 
above to find the relation 
between energy, spin and angular momentum for long and short strings.
After doing the expansion, we notice that the energy and angular 
momentum do not diverge for long strings. This is a new type of
behaviour. Even though this geometry is very 
similar to the one described in section 4.1 of the paper 
\cite{Gubser:2002tv}, we have here a warping factor.
We get that 
the relation for long strings is of the form
\be
E - J = k + \cdots,
\ee 
where $k$ is a numerical constant.

\subsection{Strings moving in a general background}

Consider now the motion of strings in backgrounds of the form,
\beq
ds_{10}^2= f(r) \left[- dt^2 + d\rho^2 + \rho^2 d\Omega_2^2(\theta,\phi)\right] + 
dr^2 +...
\eeq
where the ``...'' can be whatever one wants, the string will not be
moving in these directions. Backgrounds of this form are 
interesting since they have the general form of gravity duals to gauge 
theories in $3+1$ dimensions with a low number of supersymmetries, and
that may exhibit confinement.
Consider a string configuration that could be interpreted as a 
spinning string in the $3+1$ manifold
\beq
r= r(\sigma), \;\; t=\kappa \tau,\;\; \phi=\omega\tau,\;\; 
\rho=\rho(\sigma).
\eeq
The Polyakov action in this case is,
\beq
I= \frac{1}{2\pi \alpha'}\int d\sigma \left[r'^2 + f(r) ( \rho'^2 + \kappa^2 
- \rho^2\omega^2)\right] ,
\label{apge}
\eeq
and the constraint is,
\beq
f(r) (\rho'^2 - \kappa^2   
+ \rho^2\omega^2) + r'^2=0 .
\eeq
We can see that the derivative of the constraint will have the form 
\beq
2r' \left( r'' - \frac{1}{2}\frac{df}{dr}(\kappa^2 + \rho'^2 -\omega^2 \rho^2 )
\right) + 2\rho' \left( f(r)\rho'' + f(r) \omega^2 \rho + \frac{df}{dr} r' \rho'
\right) = 0 ,
\eeq
where terms inside the parenthesis will be the 
equations of motion derived from (\ref{apge}). Integrating the
equations of motion, we will obtain a relation between the variables, 
that we can substitute into the original action and follow the procedure in 
previous sections of the paper. This seems like an interesting
direction to investigate in the future. It seems possible that one
might obtain logarithms in these types of configurations.

\section{Conditions for supersymmetry of rotating membranes}

We do not expect our configurations to be supersymmetric given the
time dependence and the minimally supersymmetric background. However,
for completeness, we check this explicitly.

Let $\e$ generate a supersymmetry of the background spacetime metric,
i.e. it is a Killing spinor. This supersymmetry will be preserved by
the worldsheet if \cite{Hughes:fa, Bergshoeff:1987qx,Bergshoeff:1987dh}
\be\label{eq:SUSY}
\G_{M2} \e = \e ,
\ee
where
\be
\G_{M2} = \frac{1}{\sqrt{-\det \g}} \frac{1}{3!} \e^{ijk} \del_i X^\mu
\del_j X^\nu \del_k X^\s \G_{\mu\nu\s} .
\ee
Where $\G_{\mu\nu\s}$ is the standard antisymmetric combination of
eleven dimensional Dirac matrices, and $\g_{ij}$ is the induced metric
on the worldsheet.

This formula is particularly easy to apply to the membrane
configurations on $G_2$ backgrounds. For the $\B_7$ metrics, the
parallel spinor \cite{Cvetic:2001ya} has constant coefficients and
satisfies three projection conditions
\be\label{eq:projec}
\G_{2635} \e = \e ,\;\; \G_{1634} \e = \e ,\;\; \G_{6201} \e = \e ,
\ee
using tangent space indices.  The orthonormal frame on the $G_2$ is
\bea
e^0 = dr, \;\; e^1 = a (\S_1-\s_1), \;\; e^2 = a(\S_2 - \s_2), \;\; e^3 =
d (\S_3 - \s_3), \nonumber \\
e^4 = b (\S_1+\s_1), \;\; e^5 = b(\S_2 + \s_2),
\;\; e^6 = d (\S_3 + \s_3) .
\eea
From here one uses the membrane configurations of Table 1 to calculate
$\G_{M2}$. For example, for the I$_{\B}$ configuration one obtains
\be\label{eq:notcons}
\G_{{\rm I}_{\B}} = \frac{\left[\w \G_{\phi_3} + \nu_2 \G_{\phi_4} +\k
\G_t\right] \G_r\G_{\psi_3}}{c/N \sqrt{\k^2-b^2
\w^2-a^2\nu_2^2}} ,
\ee
where
\bea
\G_{\phi_3} & = &  b \sin\frac{\psi_3}{2}\G_4 + b
\cos\frac{\psi_3}{2}\G_5, \nonumber \\
\G_{\phi_4} & = & a \sin\frac{\psi_3}{2}\G_1 + a \cos\frac{\psi_3}{2}\G_2
, \nonumber \\
\G_{\psi_3} & = & \frac{c}{N} \G_6 , \nonumber \\
\G_{r} & = & \G_0 , \;\;\;\;\; \G_{t} = \G_{\hat{t}} .
\eea
The matrix (\ref{eq:notcons}) is easily seen not to commute or
anticommute with the projectors of equation (\ref{eq:projec}) and therefore no
supersymmetries are preserved. The same will be the case for the II$_{\B}$,
III$_{\B}$ and IV$_{\B}$ configurations.

The ${\Bbb D}_7$ cases are a little more complicated because the parallel
spinor \cite{Cvetic:2001kp} does not have constant
coefficients. However, it
will be sufficient for us to know that the parallel spinor satisfies
\be
\G_{2536} \e = \e ,
\ee
where we are using tangent space indices and the vielbeins are
\bea
e^0 = dr, \;\; e^1 = a (\S_1+ g\s_1), \;\; e^2 = a(\S_2 + g\s_2), \;\; e^3 =
c (\S_3 + g_3 \s_3), \nonumber \\
e^4 = b\s_1, \;\; e^5 = b\s_2,
\;\; e^6 = f \s_3 .
\eea
One then calculates $\G_{M2}$ using Table 2. For the I$_{\Bbb D}$
configuration, for example, one obtains
\be
\G_{{\rm I}_{\Bbb D}} = \frac{\left[ \w_1 \G_{\phi_1} + \nu
\G_{\phi_2} + \k \G_{t} \right] \G_r \G_{\psi_3}}{C/N \sqrt{\k^2-\w_1^2
B^2-\nu^2 A^2}} ,
\ee
where
\bea
\G_{\phi_1} & = & \frac{a g}{\sqrt{2}} \left[ (\sin\frac{\psi_3}{2} +
\cos\frac{\psi_3}{2})\G_1 + (\cos\frac{\psi_3}{2} 
-\sin\frac{\psi_3}{2})\G_2 \right] \nonumber \\
 & & + \frac{b}{\sqrt{2}} \left[
(\cos\frac{\psi_3}{2}+ \cos\frac{\psi_3}{2})\G_4 +
(\cos\frac{\psi_3}{2}-\sin\frac{\psi_3}{2}) \G_5\right] , \nonumber \\
\G_{\phi_2} & = & a (\sin\frac{\psi_3}{2}+\cos\frac{\psi_3}{2} )\G_1 +
a (\sin\frac{\psi_3}{2} - \cos\frac{\psi_3}{2})\G_2
, \nonumber \\
\G_{\psi_3} & = & \frac{c g^2}{N} \G_2 + \frac{f}{2N} \G_6 ,
\nonumber \\
\G_t & = & \G_{\hat{t}}, \;\;\;\;\;\; \G_r = \G_0 .
\eea
One can now see that $\G_{{\rm I}_{\Bbb D}}$ and $\G_{2536}$ do not
commute or anticommute and therefore no supersymmetry is preserved.
It is easy to check that the same occurs for the other configurations,
II$_{\Bbb D}$, III$_{\Bbb D}$ and IV$_{\Bbb D}$. Thus, as expected,
none of our configurations are supersymmetric.


\end{document}